%% file: 15581paper.tex
\documentclass[traditabstract]{aa}
\usepackage{graphicx}
\usepackage{txfonts}
\usepackage{booktabs}
\usepackage{natbib}
\usepackage{multirow}
\usepackage{longtable}
\usepackage[utf8]{inputenc}
\bibpunct{(}{)}{;}{a}{}{,}

\usepackage[usenames,dvipsnames]{color}
\usepackage{colortbl,xcolor}
\newcommand{\gray}{\rowcolor[gray]{1}}
\definecolor{grey}{gray}{1}

\newcolumntype{g}{l}
\newcolumntype{G}{r}
\newcolumntype{H}{c}

\newcommand{\gcn}{GCN Circ.}

\newcommand{\swift}{\textit{Swift}}

\begin{document}

\title{The circumburst density profile around GRB progenitors: 
\\ a statistical study}

\author{
S.Schulze\inst{1},
S.~Klose\inst{2},
G.~Björnsson\inst{1},
P.~Jakobsson\inst{1},
D.~A.~Kann\inst{2},
A.~Rossi\inst{2},
T.~Kr\"uhler\inst{3,4},
J.~Greiner\inst{3},
P.~Ferrero\inst{5, 6}
}

\offprints{S. Schulze, steve@raunvis.hi.is}

\institute{
   Centre for Astrophysics and Cosmology, Science Institute,  University of Iceland, Dunhagi 5, IS--107 Reykjav\'ik, Iceland 
\and
   Thüringer Landessternwarte Tautenburg, Sternwarte 5, D--07778 Tautenburg, Germany 
\and
   Max-Planck-Institut für Extraterrestrische Physik, Giessenbachstraße, D--85748 Garching, Germany 
\and
   Universe Cluster, Technische Universität München, Boltzmannstraße 2, D--85748, Garching, Germany 
\and
   Instituto de Astrof\'{\i}sica de Canarias (IAC), E--38200 La Laguna, Tenerife, Spain 
\and
   Departamento de Astrof\'{\i}sica, Universidad de La Laguna (ULL), E--38205 La Laguna, Tenerife, Spain 
}

\date{Received August 13, 2010 / Accepted October 26, 2010}
 
\authorrunning{Schulze et al.}

\abstract{According to our present understanding, long gamma-ray bursts (GRBs) originate from
the collapse of massive stars, while short bursts are caused by to the coalescence of compact
stellar  objects. Because the afterglow evolution is determined by the circumburst
density profile, $n(r)$, traversed by the fireball, it can be used to distinguish
between a constant density medium, $n(r) = \rm{const.}$, and a free stellar wind,
$n(r) \propto r^{-2}$. Our goal is to derive the most probable circumburst density
profile for a large number of \swift-detected bursts using well-sampled
afterglow light curves in the optical and X-ray bands. We combined all publicly
available optical and \swift/X-ray afterglow data from June 2005 to September 2009
to  find the best-sampled late-time afterglow light curves.  After applying several
selection criteria, our final sample consists of 27 bursts, including one short burst.
The afterglow evolution was then studied within the  framework of the fireball
model. We find that the majority (18) of the 27 afterglow light curves are compatible
with  a constant density medium (ISM case). Only 6 of the 27
afterglows show evidence of a wind profile at late times.  In particular,
we  set upper limits on the wind termination-shock radius, $R_T$, for
GRB fireballs that are propagating into an ISM profile and lower limits on
$R_T$ for those that were found to propagate through a wind medium. Observational
evidence for ISM profiles dominates in GRB afterglow studies, implying that most GRB
progenitors might have relatively small wind termination-shock radii. A smaller group of
progenitors, however, seems to be characterised by significantly more extended wind regions.}


\maketitle

\section{Introduction}

Starting with the discovery of the gamma-ray burst-supernova (GRB-SN)
association GRB 980425/SN~1998bw \citep{Galama1998a, Iwamoto1998a,
Sollerma2000a}, there is by now convincing evidence that the progenitors of
long GRBs are massive stars exploding as type Ic SNe \citep[for a review see]
[]{Woosley2006a}. Within this picture, the optical light observed after 
a long GRB is the superposition of the afterglow light, a supernova component,
and light from the underlying host galaxy (plus potential  additional radiation components at
very early times, which we will not consider here). Phenomenologically,
this immediately unveils two observing strategies to reveal a massive-star
origin of a GRB: ($i$) via the detection of a late-time SN bump (method $i$)  in the
optical light curve \citep[e.g.,][]{Reichart1999a, Galama2000a, Dado2002a,
Zeh2004a} and ($ii$) by the spectroscopic confirmation (method $ii$) of associated
supernova light \citep[the best case so far being GRB 030329:][]
{Hjorth2003a, Kawabata2003a, Matheson2003a, Stanek2003a}. In addition to both
observing strategies, there are two further methods by which a massive-star origin
can be revealed. Some afterglow spectra showed  blue-shifted
absorption line systems (method $iii$), which can be understood as signatures
from  the expanding pre-explosion wind escaping from the GRB progenitor (e.g.,
\citealt{Mirabal2003a, Schaefer2003a, Klose2004a, Starling2005a, Berger2006, 
Fox2008a, CastroTirado2010a}). Some authors, however, notice that several
properties of the putative blue-shifted absorption line systems disagree
with the expectations from Wolf-Rayet (WR) winds, e.g., line widths, ionisation levels and
metallicities \citep{Chen2007a, Prochaska2007a, Fox2008a}. Finally (method $iv$),
the circumburst medium determines the spectral and temporal evolution of the
afterglow, allowing us to discern between a constant-density medium,
$n(r)= \rm const.$, and a wind medium, $n(r)\propto r^{-2}$ \citep{Sari1998a, Chevalier1999a,
Chevalier2000a}. This method was successfully applied in, e.g., \citet{Starling2009a}
and \citet{Curran2010a}.

Naturally,  these various approaches 
have their observational advantages and disadvantages. While method $(ii)$
can provide the strongest observational evidence for a massive-star origin of the
GRB under consideration,  it can only be applied to the nearest and hence
brightest events up to a redshift of about 0.5. Even 13 years after the
first discovery of an afterglow \citep{Costa1997a, Paradijs1997a}, i.e., after more
than  500 GRBs with detected (X-ray, optical, radio) afterglow
light\footnote{http://www.mpe.mpg.de/$\sim$jcg/grbgen.html}, secure evidence for a
spectroscopically associated SN was only reported for roughly 1\% of all
events (GRBs 980425: \citealt{Galama1998a};
030329: \citealt{Hjorth2003a, Kawabata2003a,
Matheson2003a, Stanek2003a}; 031203: \citealt{Malesani2004a}; 060218:
\citealt{Ferrero2006a, Mirabal2006a, Modjaz2006a, Pian2006a, Sollerman2006a}; 081007:
\citealt{DellaValle2008a}; 100316D: \citealt{Chornock2010a, Starling2010a}). Contrary to this
approach, method $(i)$ can basically reveal a SN component up to a redshift
of about 1 \citep{Zeh2004a}, assuming a non-extinguished SN 1998bw as a template (the most distant
SN bump was found for GRB 000911 at $z=1.06$; \citealt{Masetti2005a}). At notably
higher redshifts a GRB-SN becomes too faint to be discovered even
with 8m-class optical telescopes because of line blanketing \citep[e.g.,][]{Filippenko1997a}.

In principle, methods $(iii)$ and $(iv)$ do not have redshift constraints,
because both rely on the observation of the afterglow and not on the (expected)
SN component.  In addition, method $(iv)$ splits into different approaches and
basically works for the optical and X-ray band in the same way.  Moreover,
it can even be applied to the most distant GRBs, which are already affected by
Lyman dropout in the optical bands.  Its main disadvantage is that
it usually requires substantial observational efforts. 
In particular, it relies on the measurement of the light-curve
evolution, including the determination of the spectral energy distribution
(SED).

In this paper, applying method ($iv$), we use late-time data of afterglows
detected by \swift\ to tackle the question of the preferred density
profiles in a statistical sense. Our goal is to combine all  publicly
available optical data with \swift/XRT data in order to determine the
corresponding circumburst density profile, $n(r)\propto
r^{-k}$. Qualitatively, this splits into either a constant-density medium (hereafter referred to as the interstellar medium or ISM)
($k=0$) or a free wind profile ($k=2$).  While the complex wind history of an
evolved massive star might produce density profiles different from the ideal
case ($k=0, 2$; \citealt{Crowther2007a}), in general the data do not allow for a more accurate
determination of $k$, but only to distinguish between these two cases.

It should be stressed that the intention of our study is not to provide ultimate
conclusions on the density profiles found for individual bursts. Instead, it
is meant as a statistical approach using bursts with the best available X-ray
as well as optical data. The questions we want to address are: (1) What is, in
a  statistical sense, the  preferred  circumburst density profile? (2) What
is the ratio between events with ISM and with wind profiles? (3)
What does this tell us about the typical radius of the wind  termination shock
that is expected to exist in the ambient medium  surrounding a long burst GRB
progenitor?

Throughout the paper we use the convention $F_\nu \left(t\right) \propto t^{-\alpha}\nu^{-\beta}$
for the flux density, where $\alpha$ is the temporal slope and $\beta$ is the
spectral slope. All errors are $1\sigma$ uncertainties unless noted otherwise.

\section{Data selection}\label{sec:data}
\subsection{Data gathering}\label{sec:gathering}

The optical data were taken from a photometric database maintained and updated
by one of the co-authors (D.A.K.). In addition, we added data for GRB 090726 from
\citet{Simon2010a}, \citet{Fatkhullin2009a}, \citet{Haislip2009a},
\citet{Kelemen2009a}, \citet{Landsman2009a}, \citet{Sakamoto2009a} and
\citet{Volnova2009a}. The properties of the optical afterglow sample, the data
gathering and the deduced SEDs of the afterglows are discussed in 
\cite{Kann2006a, Kann2008a, Kann2010a}. Furthermore, we compared their SED results
with the work of \cite{Schady2007a, Schady2010a}. The latter authors added X-ray data
to model the spectral energy distributions from roughly 1 eV to 10 keV.
Finally, to create a denser light-curve coverage, all non-$R_C$ band data in
each light curve were shifted to the $R_C$ band. In doing this, we used the
colours of the SED, assuming no spectral evolution and omitting data where
clear colour evolution was evident.

The X-ray data were retrieved from the \swift~data archive and the light curves
from the \swift~light-curve repository (version February 2010) updated and maintained by
\citet{Evans2007a, Evans2009a}. Following \citet{Nousek2006a}, we reduced the data with
the software package \texttt{HeaSoft} 6.6.1\footnote{
\texttt{http://heasarc.gsfc.nasa.gov/docs/software/lheasoft}} together with
the calibration file version \texttt{v011}\footnote{\texttt{heasarc.gsfc.nasa.gov/docs/heasarc/caldb/swift}}.
Furthermore, we applied the methods detailed in \citet{Moretti2005a}, \citet{Romano2006a}, and
\citet{Vaughan2006a} to reduce pile-up affected data. We extracted SEDs
at different epochs with approximately 500 background subtracted counts, to check for
spectral evolution. If the properties of the SED (spectral slope and absorption) did
not evolve, we extracted a new SED from the maximum possible time interval. In addition, we
used the Tübingen absorption model by \cite{Wilms2000a} and their interstellar-medium
metal abundance template. The Galactic absorption was fixed to the weighted mean based on
\cite{Kalberla2005a}. We included the Chandra light curve of GRB 051221A
from \citet{Burrows2006a}.

\subsection{Sample definition and light-curve fitting}\label{sec:sample}

Among all \swift~GRBs observed until September 2009, we selected from our
photometric database, which contains long and short bursts, those
90 bursts that have an optical and an X-ray afterglow as well as a measured
spectroscopic or a photometric redshift. From these we selected those bursts
with the best-sampled optical and X-ray afterglow light curves as follows.

First, we required  information on the spectral slope of the afterglow in the X-ray band. We
derived this information from publicly available data. In the optical bands,
however,  multi-band data are usually
not available. In these cases  we used the results from \cite{Kann2008a, Kann2010a}
and \cite{Schady2007a, Schady2010a}, with the latter being based
on optical-to-X-ray SED fits.

\input{15581tpred.tex}

Second, we required that an afterglow light curve can be fitted with a
multiply broken power-law and is not dominated by flares or bad data
sampling. After excluding time intervals affected by flares, we fitted
the light curves with a smoothly broken power-law of the order $m$ (see Appendix \ref{app:mpl}
for its definition) with a Simplex and a Levenberg-Marquardt algorithm \citep{Press2007a}. 
Furthermore, we transformed $R_c$-band light curves to flux densities
with the zero-point definition of \citet{Bessell1979a}. For the X-ray regime
we followed \cite{Gehrels2008a}; the flux density, $F_{\nu,\rm x}$, in $\mu$Jy
at the frequency $\nu_{\rm x}$ is then given by 
\begin{eqnarray*}
\label{eq:meanfrequency} 
F_{\nu,\rm x} = 4.13\times10^{11}\ \frac{(1-\beta_{\rm x})F_{\rm x}}
{(10\,\rm{keV})^{1-\beta_{\rm x}}-
(0.3\,\rm{keV})^{1-\beta_{\rm x}}}\ E_{\rm x}^{-\beta_{\rm x}}\,,
\end{eqnarray*}
where $\beta_{\rm x}$ is the spectral slope in the X-ray band,
$F_{\rm x}$ is the measured flux in the 0.3--10 keV range in
units of erg\,cm$^{-2}$\,s$^{-1}$ and the reference energy $E_{\rm x}$
is given in keV. For all bursts we chose the logarithmic mean between
0.3 keV and 10 keV as a reference, i.e., $E_{\rm x}$=1.73 keV ($\nu_{\rm x}=4.19\times10^{17}\,\rm{Hz}$).
The numerical constant converts the energy in units of keV to a frequency
in units of Hz and the flux density from erg\,cm$^{-2}$\,s$^{-1}$\,Hz$^{-1}$
to $\mu$Jy.

Third, once power-law segments had been defined in the afterglow light curves,
we excluded from further studies those bursts where the difference in the
late-time decay slopes between the optical and the X-ray band could not be
explained within the framework of the fireball model (Table \ref{tab:predictions}); i.e., the difference
in decay slopes, $\alpha_{\rm x} - \alpha_{\rm opt}$, was larger than 1/4 within $3\sigma$.
Because of this, a strict criterion for the beginning of  the late-time
evolution of an afterglow cannot be given. Evidence for  an \textit{observed}
canonical light curve does not exist in the optical \citep{Kann2008a, Kann2010a,
Panaitescu2010a}, but may exist in the X-rays \citep[e.g.,][]{Nousek2006a, Zhang2006a, Evans2009a}. 
Therefore, as an operational definition an afterglow is in its late-time phase
when its temporal and spectral evolution can be explained by the fireball model
from a particular time after the corresponding GRB.

After applying these selection criteria, about half of the 90 bursts had to
be rejected owing to bad sampling, poor data quality, flares, and other peculiarities.
Another quarter had to be rejected because of $|\alpha_{\rm x} - \alpha_{\rm opt}|> 1/4$
within $3\sigma$. All data were then corrected for host extinction in the optical, $A^{\rm host}_{\rm V}$,
if needed, and for Galactic and host absorption in the X-ray band,
$N^{\rm host} _{\rm H}$.

In total, 27 afterglows passed these selection criteria.
The sample consists of 25 long and one short GRB (GRB 051221A), and one controversial
event (GRB 060614) in terms of the short/long classification scheme.
Their input data are summarised in Table \ref{tab:sed} and \ref{tab:input}.

\section{Results and discussion}

\subsection{The circumburst medium density profile}
\label{sec:fits}\label{sec:fits_ensemble}\label{sec:cbm}
\subsubsection{Identifying the circumburst medium}\label{sec:indentifying}
As a first step in the identification of the circumburst medium, we defined
nine possible spectral and dynamical regimes  (Table \ref{tab:closure}). The
spectral regimes separate according to the position of the cooling frequency
with respect to the observer frame, while the dynamical regimes distinguish
between a spherical and a jetted evolution.

\input{15581tclo.tex}

To derive the most probable density profile into which a GRB jet propagates,
we proceeded in the following way. The main criterion was that the result agrees
with the optical as well as with the X-ray data.  In doing so, we
first analysed the closure relations for the nine models (Table
\ref{tab:closure}) in the optical and X-ray bands and selected only those
relations (models) which were fulfilled within $3\sigma$. Second, we computed
the difference in the decay slopes,  $ \alpha_{\rm x} -
\alpha_{\rm opt}$, and distinguished between the models according to the three
possible cases $-1/4, 0, +1/4$ (Table \ref{tab:predictions}), again within
$3\sigma$.   Third,  if possible, we took into account the difference in the
spectral slope,  $\beta_{\rm x} - \beta_{\rm opt}$, which is
either 0 or 1/2 (see Table \ref{tab:predictions}). If this criterion could be
applied, we required that it is fulfilled within $3\sigma$.
In addition we required that the $1\sigma$ uncertainty in the spectral
or temporal decay slopes was less than 0.2.

The final results of the light curve fits of the 27 bursts considered
are presented in Fig. \ref{fig:sample}. In addition, we plot here the observed
flux-density ratio $\left(F_{\nu, \rm opt}/F_{\nu, \rm x}\right)(t)$ (middle panels in Fig.~\ref{fig:sample})
as a function of time. According to Table \ref{tab:predictions}, the expected flux-density ratio
is only allowed to take a certain value depending on the spectral and dynamical
regime. We checked if the observed flux-density ratio,  $\left(F_{\nu, \rm opt}/F_{\nu, \rm x}\right)(t)$
(middle panels in Fig.~\ref{fig:sample}), agreed with the model(s)
that successfully passed the previous three criteria. The allowed parameter space of
the flux-density ratio of all considered models (Table \ref{tab:predictions}) is shown as grey box in the
middle panels in Fig.~\ref{fig:sample}. The upper and lower boundary always refer to
$\nu_c \leq \nu_{\rm opt}$ with $F_{\nu, \rm opt}/F_{\nu, \rm x} = (\nu_{\rm
  opt}/\nu_{\rm x})^{-p/2}$ and to  $\nu_c \geq \nu_{\rm x}$ with $F_{\nu, \rm
  opt}/ F_{\nu, \rm x} = (\nu_{\rm opt}/\nu_{\rm x})^{-(p-1)/2}$.
This criterion came into play when we were unable to distinguish
whether the cooling frequency was redward of the optical band or blueward of
the X-ray band, in other words when the flux-density ratio agreed either
with the lower or upper boundary in Fig. \ref{fig:sample}.

\subsubsection{Ensemble properties}\label{sec:ensemble}

Combining these different criteria, we found that for about 60\% (16/27) of all studied
cases the cooling break was between the optical and the X-ray band (Table
\ref{tab:cbm}), i.e., the difference in the decay slope  is either $+1/4$ or
$-1/4$ (for a constant density profile and a free wind medium, respectively).
In total, we could identify the circumburst density profile  (ISM or wind) for
25 of the 27 investigated afterglows (Table \ref{tab:cbm}). However, we
identified a wind medium for only six events (GRBs, 050603, 070411, 080319B,
080514B, 080916C and 090323). The other 19 bursts were consistent with an ISM
profile except for GRBs 051221A and 060904B.

Our procedure to find physical descriptions of light-curve segments, in other words
identifying the spectral regime and the circumburst medium, allowed us to find 
descriptions of more optical and X-ray afterglows than are presented in the
literature (Table \ref{tab:cbm}). Our results usually agree with the literature
(for references see Table \ref{tab:cbm}); they only differ for GRBs 070802, 080721,
090323, and 090328. The contradiction in the latter three bursts is due to the size
of the optical data set. We used the maximum publicly available data set in contrast
to \cite{Starling2009a} (GRB 080721) and \cite{Cenko2010a} (GRBs 090323 and 090328).

Four bursts are of particular interest in our sample: 
(i) GRB 051221A ($z=0.546$; \citealt{Soderberg2006a}) is a short burst \citep{Burrows2006a},
i.e., most likely it originated from the merger of two compact objects
\citep{Blinnikov1984a, Paczynski1986a, Goodman1986a, Eichler1989a}.
Unfortunately, we cannot discern between a wind or an ISM profile because $\nu_c<\nu_{\rm x}$.  
The cooling frequency was below the optical bands so that neither the optical nor
the X-ray data can be used to reveal the circumburst medium except during a post-jet
break phase without lateral spreading.

(ii) GRB 060614 ($z=0.125$; \citealt{DellaValle2006b}) is a quite 
controversial event that does not easily fit into the classical short/long
classification scheme \citep{DellaValle2006a, Fynbo2006a, GalYam2006a,
  Gehrels2006a, Mangano2007a, Zhang2007a, Kann2008a, Zhang2009a}. 
The difference in the spectral slope, $\beta_{\rm x} - \beta_{\rm opt} = 0.00 \pm 0.11$
(Table \ref{tab:cbm}), rules out that the cooling frequency lies between
the optical and X-ray bands. Furthermore, we did not find evidence
for a wind medium. The optical and X-ray data exclude a wind
medium with high confidence. The deviation between the observed and
predicted temporal decay slope is $<1.3\sigma$ and $>3.5\sigma$ for an ISM
and a wind medium, respectively. In our sample this burst
belongs to a small number of cases where the flux density ratio could be used
to distinguish between $\nu_c > (\nu_{\rm opt}, \nu_{\rm x})$ and $\nu_c <
(\nu_{\rm opt}, \nu_{\rm x})$. The small observed flux density ratio agrees
with $\nu_c > (\nu_{\rm opt}, \nu_{\rm x})$.

(iii) GRB 060904B has a well-defined light curve  (Fig. \ref{fig:sample}, Table \ref{tab:input}) and SED
(Table \ref{tab:sed}) in the optical and X-ray bands, respectively. The difference in the decay
slopes, $\alpha_{\rm x} - \alpha_{\rm opt} = 0.20\pm0.04$ (Table \ref{tab:cbm}), favours an ISM
profile with the cooling frequency lying between the optical and the X-ray bands. The difference in the spectral slopes,
however, does not support this scenario, $\beta_{\rm x} - \beta_{\rm opt} = 0.00\pm0.16$ (Table
\ref{tab:cbm}). Rather, both afterglow components seem to be in the same spectral regime. Therefore,
we could not find a consistent description of the optical and X-ray afterglow. The main reason could
be that the optical data are mainly based on preliminary data (see \citealt{Kann2010a} for
details on the data gathering).

(iv) GRB 080319B ($z=0.937$;
\citealt{DElia2009a}) is the only burst in our sample with a photometrically
detected supernova component \citep{Bloom2009,Tanvir2008a}. 
An ISM profile can be ruled out with very high confidence. The difference
in the spectral slopes, $\beta_{\rm x} - \beta_{\rm opt} = 0.48 \pm 0.12$
(Table \ref{tab:cbm}), favours the cooling break to be in between
the optical and the X-ray bands. The deviation between $\Delta \alpha _{\rm obs}= \alpha_{\rm x} - \alpha_{\rm opt}$ and $\Delta\alpha_{\rm predicted}$ is
$1\sigma$ for a wind medium but $9\sigma$ for an ISM profile. The closure
relations support this finding. The deviation between the observed and predicted
optical decay slope is $0.1\sigma$ for a wind medium and $4.6\sigma$ for an ISM profile
(see also \citealt{Racusin2008a}). 

\subsubsection{The electron index}

Finally, the identification of the light-curve segments allowed us to derive the
electron index, $p$. It is shown in the top panels in Fig. \ref{fig:sample} for every
burst. In agreement with other studies \citep[e.g.,][]{Panaitescu2001a, Shen2006a, Zeh2006a, Starling2008a,
Curran2009b, Ghisellini2009a, Curran2010a} we find that the distribution extends from $p\sim2$ to
$p\sim3$, indicating that $p$ is no universal value.

\subsection{Constraining the wind - ISM transition zone}

Once we had identified the circumburst density profiles, we could investigate if
there is observational evidence for the position of
the wind termination-shock radius, where the density profile changes from that
of a free wind ($k=2)$ to a density profile with $k=0$ (independent of whether
it is the shocked wind or the ISM; see \citealt{Peer2006a,
  vanMarle2007a}). While the theory of a blastwave crossing such a density
discontinuity has been worked out \citep{Peer2006a}, finding the corresponding
observational signature is difficult.  Even though data on several hundred
afterglows exist, they do not provide this kind of information in a convincing
way \citep{Starling2008a, Curran2009b}.  This leaves open the question of
observationally determining the wind termination-shock  radius.

Here we cannot determine the radius of the wind termination shock
for any (long)  GRB progenitor either. However, we can characterise its
position in a statistical sense. The light curves in Fig. \ref{fig:sample} show
$F_{\rm opt}/F_{\rm x}$ as a function of time. Its first logarithmic derivative,
$\left(\alpha_{\rm x}-\alpha_{\rm opt}\right)\left(t\right)$, reveals either the maximum
time up to which a wind profile is identified or it reveals the minimum time
after which an ISM profile agrees with the data. The function
$\left(\alpha_{\rm x} - \alpha_{\rm  opt}\right)(t)$ is a smooth function in
time in contrast to the fit values, because we used a smoothly broken power
law to describe the light-curve evolution. Table \ref{tab:radii} summarises the time
intervals for which the asymptotic values for $\alpha_{\rm x} - \alpha_{\rm opt}$
were reached, i.e. $-$1/4, 0, or 1/4 within $1\sigma$.

\input{15581trad.tex}

In the observer frame, the radius of the fireball propagating into a free
wind medium is \citep{Chevalier2000a} 
\begin{equation}
R(t) = 1.1\,\times\,10^{17}\ \left(\frac{2t\, 
E_{\rm{iso}, 52}}{(1+z)\,A_\star} \right)^{1/2} \ \ {\rm cm}\,,
\label{Radii}
\end{equation}
where $t$ is measured in units of days, $E_{\rm iso}$ is the
isotropic equivalent energy in units of $10^{52}\,\rm{erg}$ and $A_\star$ is defined via $A=\dot M_{\rm{w}}/4\pi v_{\rm{w}}=5\,\times\,10^{11}\,A_\star\, \rm{g}\, \rm{cm}^{-1}$,
with $\dot M_{\rm w}$ being the mass-loss rate, and $v_{\rm w}$
the wind velocity. The quantity $A_\star$ refers to a mass-loss rate
of $\dot{\rm{M}}_{\rm{w}} = 10^{-5}\,M_{\sun}\,\rm{yr}^{-1}$ and a velocity of the stellar wind
of $\rm{v}_{\rm{w}} =10^8\,\rm{cm}\,s^{-1}$.

Using the aforementioned approach and fixing for simplicity  $A_\star=1$,
Table \ref{tab:radii} summarises the deduced upper and lower limits on the wind
termination-shock radius.  The distribution extends over three orders of
magnitude (from $\approx10^{-3}$ pc to 1 pc).  This spread is partly due to
the strong correlation with $E_{\rm iso}$ ($R(t)\propto E^{1/2} _{\rm iso}$)
and is thus related to the spread in the energy released during the prompt
emission in gamma-rays  (width $\approx 4.3$ dex).
On the other hand, since $R(t)$ scales with $A^{-1/2} _\star$, 
$A _\star$ would have
to vary by a factor of 100 in the right way to reduce the width 
of this distribution by only a factor of 10.

\begin{figure}
\includegraphics[bb=40 17 327 274, clip, angle=0, width=0.49\textwidth]{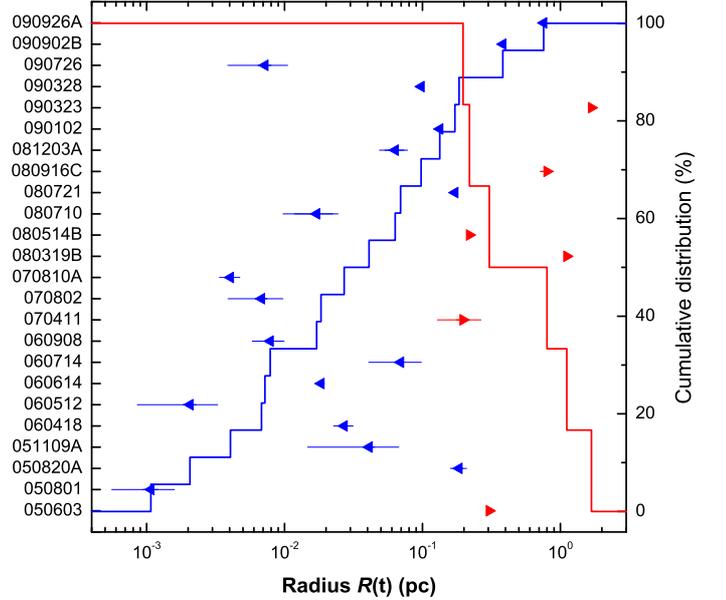}
\caption{Shown here are lower (to-the-right pointing triangles) and upper
(to-the-left pointing triangles) limits on the position of the wind
termination shock based on Eq.~\ref{Radii}, assuming $A_\star=1$ in all cases (Table \ref{tab:radii}). Note that
GRB 060614 is a much debated burst (see Sect.~\ref{sec:indentifying}). The step curves are the cumulative
distributions to the lower limits up to which a wind profile is identified
and upper limits after which a constant density medium (ISM) agrees with the data.}
\label{fig:radii}
\end{figure}

Is the width of the distribution we have found for the upper limits on the wind termination shock
reasonable?  Based on numerical wind models of Wolf-Rayet stars, \cite{Fryer2006a} found that for
$A_\star=1$ the radius of the termination shock is approximately given by
\begin{equation}
 R_{\rm T}= (n_{\rm ISM}/100\ \mbox{\rm cm}^{-3})^{-1/2}\ \mbox{\rm pc}\,.
\end{equation}
ISM densities of the order of $10^6$ cm$^{-3}$ are then required to reduce $R_T$ to 0.01 pc.
These large-scale gas densities are rather unique and only typical for dense cores of molecular clouds
\citep[with the Rho Ophiuchi Cloud as an example, e.g.,][]{Klose1986a}.

Observationally, gas densities could only be derived for a few bursts because of the lack of radio data.
So far, the highest values are about 600 cm$^{-3}$ \citep{Frail2006a, Thoene2008a}, while the nominal
value is about a factor of 100 smaller \citep{Frail2006a}. These measurements do not necessarily
rule out the model of \citet{Fryer2006a}. Successful radio observations might have picked out a certain
class of GRBs. Furthermore, radio observations are very challenging at early times because the brightness
of the afterglow is increasing in the radio bands while the afterglow is already decaying in the optical
and X-ray bands \citep{Zhang2004a}. Thus, it is difficult to extract information on the direct vicinity
of the progenitor. On the other hand, if the average particle density of the circumburst medium is about
1--10 cm$^{-3}$, then additional mechanisms are required to bring the wind termination-shock
radius closer to the star \citep[e.g.,][]{vanMarle2006a}.

The other possibility to reduce $R_T$ is to decrease $A_\star$, since $R_T \propto A_\star^{1/2}$
\citep{Chevalier2004a}. For example, for $A_\star=0.01$ all data points in Fig. \ref{fig:radii}
would shift along the $R$-coordinate to higher values by a factor of 10 (Eq. \ref{Radii}), while
$R_{\rm T}$ would decrease by the same factor. In this case, lower circumburst gas densities
would be required. This touches upon the question on how small $A_\star$ can be. Studies of WR stars
do not favour values of less than 0.01 in polar directions \citep[][his Table 1]{Eldridge2007a}.
Moreover, observations of nearby WR stars do not show evidence for these low values either
\citep{Nugis2000a}. However, the majority of nearby WR stars are surely not seen pole-on, in contrast
to GRBs.  Therefore, it is difficult to decide if these observational constraints 
on WR stars can be applied to GRB progenitors.

On the other hand, the six GRBs that favour a free wind medium (Table \ref{tab:cbm}) have large
lower limits on the wind termination-shock radius (Table \ref{tab:radii}). This matches theoretical
models by \citet{vanMarle2007a, vanMarle2008a}, which allow $R_{\rm T}$ to extend up to several parsecs.

The separation between the lower and upper limits for wind and ISM-profiles, respectively, on the wind
termination-shock radius could be even larger. Refining the lower and upper limits is
difficult, however. The lower limits, which are deduced from $\tilde{t}_{\rm end}$ (Fig. \ref{fig:sample},
Table \ref{tab:radii}),
depend on the observing strategies due to the brightness of the afterglow and the brightness of the
underlying host galaxy in the optical bands. On the other hand, the upper limits, which are deduced from
$\tilde{t}_{\rm start}$ (Fig. \ref{fig:sample}, Table \ref{tab:radii}), can be affected by additional radiation components
at early times.

\section{Summary and conclusion}\label{sec:summary}

After applying several selection criteria   (closure relations, the
differences in the spectral and temporal slopes, and the flux density ratio, $F_{\rm opt}/F_{\rm x}$) we
selected the best-sampled \swift~GRBs with well-observed optical as well as
X-ray afterglow data from June 2005 to September 2009. Altogether 27 bursts entered our
sample, which was used to investigate the  density profile of the circumburst
medium  (constant density medium or free wind), including one short
burst (GRB 051221A) and one controversial event in terms of classification
(GRB 060614), which successfully passed our selection criteria among
all bursts. The other 25 events are classified as long bursts without doubt.

Combining optical with X-ray data is advantageous because optical data usually allow
for a more precise determination of the temporal decay slope of an afterglow,
while X-ray data can in general be used to extract the SED.
Combining both emission components substantially improves our capability to  distinguish
between an ISM and a wind medium. Thereby, we concentrated on the
late-time evolution, i.e., times when the proper afterglow is not affected
anymore by flares and additional radiation components (e.g., the reverse shock,
central engine activity). 

Our study shows that only six of the 25 long bursts (24\%) investigated here
(GRBs 050603, 070411, 080319B, 080514B, 080916C and 090323) showed evidence
for a free wind medium at late times. In the other cases (76\%),
except for the short burst GRB 051221A and 060904B, the blastwaves were propagating
into a constant density-medium. In particular, the controversial burst GRB 060614
favours an ISM profile. This is not in disagreement with a massive-star origin
as our result for long bursts indicates. 

In addition, we were able to set limits on the wind termination-shock radii
of the corresponding GRB progenitors. Only 24 of 27 bursts (Table \ref{tab:radii})
had good enough data to perform this analysis. Fixing the relative mass-loss rate to $A_\star = 1$,
the distribution we deduced covers three orders of magnitude. We find a tentative
grouping into (long) GRB progenitors with comparably small and comparably large
termination-shock radii. Whether this points to two distinct populations of (long)
GRB progenitors or if this is a selection effect, remains an open issue.
At least theoretically it is well possible that the long burst population splits
into  single star progenitors and those belonging to a binary system \citep{Georgy2009}.
Further observational data are required to reveal a potential binary nature of 
the long burst progenitors.

\section*{Acknowledgments}

We thank the referee for a very careful reading of the
manuscript and a rapid reply. S.S. acknowledges support by a Grant of Excellence
from the Icelandic Research Fund and Thüringer Landessternwarte Tautenburg,
Germany, where part of  this study was performed. D.A.K.
acknowledges support from grant DFG Kl 766/16-1. A.R. acknowledges
support from the BLANCEFLOR Boncompagni-Ludovisi, née Bildt foundation.
T.K. acknowledges support by the DFG cluster of excellence 'Origin and
Structure of the Universe'. S.S. acknowledges Robert Chapman (U Iceland),
Elisabetta Maiorano (CNR Bologna), Andrea Mehner (U Minnesota), Kim Page
(U Leicester), Eliana Palazzi (CNR Bologna) and Gunnar Stefansson (U Iceland)
for helpful discussions. This work made use of data supplied by the UK Swift
Science Data Centre at the University of Leicester.

\bibliographystyle{aa}

\appendix
\input{15581ampl.tex}
\section{Tables}
\onecolumn
\input{15581tbeta.tex}
\input{15581talfa.tex}
\input{15581tres.tex}
\twocolumn
\section{Figures of the light curve fits}
\input{15581fig.tex}

\end{document}

%% file: 15581tpred.tex
\begin{table}
\caption{Summary of the considered afterglow models.}
\begin{center}
\begin{tabular}{cccl}
\toprule
\multirow{1}{*}{Spectral regime} &$\beta_{\rm x}-\beta_{\rm opt}$&$\alpha_{\rm x}-\alpha_{\rm opt}$& \multicolumn{1}{c}{\multirow{1}*{$F_{\nu,\rm{opt}}/F_{\nu,\rm{x}}$}}      \\
\midrule
\multicolumn{4}{l}{\bf Spherical expansion}\\
\midrule
\gray	$\nu_{\rm c} < \nu_{\rm opt} < \nu_{\rm x}$  & 0		& 0			& $\left(\nu_{\rm opt}/\nu_{\rm x}\right)^{-p/2}$\\
		$\nu_{\rm opt} < \nu_{\rm c} < \nu_{\rm x}$  &$1/2$		& $\pm 1/4$	& $\nu^{-(p-1)/2} _{\rm opt}  \nu^{p/2} _{\rm x} \nu^{-1/2} _{\rm c} (t)$\\
\gray 	$\nu_{\rm opt} < \nu_{\rm x} < \nu_{\rm c}$  & 0      	& 0         & $\left(\nu_{\rm opt}/\nu_{\rm x}\right)^{-(p-1)/2}$\\
\midrule
\multicolumn{4}{l}{\bf Jet with sideways expansion}\\
\midrule
\gray	$\nu_{\rm c} < \nu_{\rm opt} < \nu_{\rm x}$  & 0 	 	& 0         & $\left(\nu_{\rm opt}/\nu_{\rm x}\right)^{-p/2}$\\
		$\nu_{\rm opt} < \nu_{\rm c} < \nu_{\rm x}$  & $1/2$  	& 0         & $\nu^{-(p-1)/2} _{\rm opt}  \nu^{p/2} _{\rm x} \nu^{-1/2} _{\rm c}$\\
\gray 	$\nu_{\rm opt} < \nu_{\rm x} < \nu_{\rm c}$  & 0      	& 0         & $\left(\nu_{\rm opt}/\nu_{\rm x}\right)^{-(p-1)/2}$\\
\midrule
\multicolumn{4}{l}{\bf Jet without sideways expansion}\\
\midrule
\gray	$\nu_{\rm c} < \nu_{\rm opt} < \nu_{\rm x}$  & 0    	& 0         & $\left(\nu_{\rm opt}/\nu_{\rm x}\right)^{-p/2}$\\
		$\nu_{\rm opt} < \nu_{\rm c} < \nu_{\rm x}$  & $1/2$	& $\pm 1/4$	& $\nu^{-(p-1)/2} _{\rm opt}  \nu^{p/2} _{\rm x} \nu^{-1/2} _{\rm c} (t)$\\
\gray	$\nu_{\rm opt} < \nu_{\rm x} < \nu_{\rm c}$  & 0      	& 0         & $\left(\nu_{\rm opt}/\nu_{\rm x}\right)^{-(p-1)/2}$\\
\bottomrule
\end{tabular}
\tablefoot{
Theoretical differences in the spectral and temporal slopes as well as
the flux-density ratio depend on the position of the cooling frequency, $\nu_c$, with respect to
the optical and the X-ray band ($\nu_{\rm opt}, \nu_{\rm x},$ respectively; 
valid for an electron index, $p$, of larger than 2;
e.g., \citealt{Zhang2004a}, \citealt{Panaitescu2007a}). Depending on the
circumburst medium the positive (ISM) or negative (wind) solution for
$\alpha_{\rm x}-\alpha_{\rm opt}$ applies.
}
\label{tab:predictions}
\end{center}
\end{table}

%% file: 15581tclo.tex
\begin{table}
\caption{The closure relations combining the temporal decay slope
$\alpha$ and the spectral slope $\beta$
	(adopted from \citealt{Zhang2004a} and \citealt{Panaitescu2007a}; valid for $p>2$).}
\centering
\begin{tabular}{l cc}
\toprule
Spectral		& \multicolumn{2}{c}{Closure relation $\alpha\left(\beta\right)$}\\
regime			& \multicolumn{1}{c}{ISM}	& \multicolumn{1}{c}{Wind}	\\
\midrule
\multicolumn{3}{l}{\textbf{Spherical expansion}}\\
\midrule
\gray $\nu_c > \nu$	& $3\beta/2$~[S1a]	&  $(3\beta+1)/2$~[S1b]\\
      $\nu_c < \nu$	& \multicolumn{2}{c}{$(3\beta-1)/2$~[S2]}		\\
\midrule
\multicolumn{3}{l}{\textbf{Jet with sideways expansion}}\\
\midrule
\gray $\nu_c > \nu$	& \multicolumn{2}{H}{$2\beta+1$~[J1]}			\\
      $\nu_c < \nu$	& \multicolumn{2}{c}{$2\beta$~[J2]}			\\
\midrule
\multicolumn{3}{l}{\textbf{Jet without sideways expansion}}\\
\midrule
\gray $\nu_c > \nu$	& $(6\beta+3)/4$~[j1a]	& $(3\beta+2)/2$ [j1b]	\\
      $\nu_c < \nu$	& $(6\beta+1)/4$~[j2a]	& $(\beta+5)/4$ [j2b]	\\
\bottomrule
 \end{tabular}
\tablefoot{Abbreviation used here for a 
certain model is given in brackets 
	(adopted from \citealt{Panaitescu2007a}; (S, J, j)=(spherical expansion, jet
	with sideways expansion, jet without lateral spreading),
	(1, 2)=($\nu<\nu_c$, $\nu>\nu_c$), (a, b)=(ISM, wind)). Entries
	extending over two columns are valid for both media.
}
\label{tab:closure}
\end{table}

%% file: 15581trad.tex
\begin{table}
\caption{Constraints on the termination-shock radii.}
\centering
\begin{tabular}{l rr l l}
\toprule
\multicolumn{1}{l}{\multirow{2}*{GRB}}	& \multicolumn{1}{c}{$\tilde{t}_{\rm start}$}	& \multicolumn{1}{c}{$\tilde{t}_{\rm end}$} 					& \multicolumn{1}{c}{$\log E_{\rm iso}$}	& \multicolumn{1}{c}{$R_T$}\\
										& \multicolumn{1}{c}{ (ks)}		& \multicolumn{1}{c}{ (ks)}	&\multicolumn{1}{c}{$\left[\rm{erg}\right]$}	& \multicolumn{1}{c}{(pc)}\\
\midrule
\multicolumn{5}{l}{\textbf{Constant density medium ($n(r)\propto r^0$)}}\\
\midrule
\gray	050801	& $0.31$	& $105.70$	& $51.51^{+0.34} _{-0.12}$	& $<0.0011^{+0.0005} _{-0.0001}$\\ 
	050820A	& $42.71$	& $411.00$	& $53.99^{+0.11} _{-0.06}$	& $<0.18^{+0.02} _{-0.01}$\\ 
\gray	051109A	& $25.87$	& $264.60$	& $52.87^{+0.08} _{-0.89}$	& $<0.04^{+0.01} _{-0.03}$\\ 
	060418	& $4.40 $	& $535.80$	& $53.15^{+0.13} _{-0.10}$	& $<0.027^{+0.004} _{-0.003}$\\ 
\gray	060512	& $10.50$	& $287.50$	& $50.30^{+0.40} _{-0.10}$	& $<0.0021^{+0.0012} _{-0.0002}$\\ 
	060614	& $51.59$	& $1497.00$	& $51.40^{+0.05} _{-0.04}$	& $<0.018\pm0.001$\\ 
\gray	060714	& $78.49$	& $286.10$	& $52.89^{+0.30} _{-0.05}$	& $<0.069^{+0.029} _{-0.004}$\\ 
	060908	& $0.92$	& $82.14$	& $52.82\pm0.20$		& $<0.008\pm0.002$\\ 
\gray	070802	& $10.90$	& $88.78$	& $51.70^{+0.31} _{-0.09}$	& $<0.007^{+0.003} _{-0.001}$\\ 
	070810A	& $1.95 $	& $34.41$	& $51.96^{+0.05} _{-0.16}$	& $<0.0041^{+0.0002} _{-0.0007}$\\ 
\gray	080710	& $23.10$	& $348.50$	& $51.90^{+0.31} _{-0.32}$	& $<0.02\pm0.01$\\ 
	080721	& $29.46$	& $1258.00$	& $54.09^{+0.03} _{-0.04}$	& $<0.17\pm0.01$\\ 
\gray	081203A	& $12.07$	& $301.00$	& $53.54^{+0.18} _{-0.15}$	& $<0.06\pm0.01$\\ 
	090102	& $77.57$	& $262.80$	& $53.30 \pm 0.04$		& $<0.13\pm0.01$\\ 
\gray	090328	& $57.89$	& $923.30$	& $52.99 \pm 0.01$		& $<0.098\pm0.001$\\ 
	090726	& $3.61 $	& $10.00$	& $52.26^{+0.33} _{-0.10}$	& $<0.007^{+0.003} _{-0.001}$\\ 
\gray	090902B	& $45.64$	& $1168.00$	& $54.49 \pm 0.01$		& $<0.383\pm0.004$\\ 
	090926A	& $325.80$	& $1798.00$	& $54.27 \pm 0.02$		& $<0.76\pm0.02$\\ 
\midrule
\multicolumn{5}{l}{\textbf{Free stellar wind ($n(r)\propto r^{-2}$)}}\\
\midrule
\gray	050603	& $39.21$	& $196.50$	& $53.79\pm0.01$		& $>0.305\pm 0.004$\\ 
	070411	& $92.09$	& $522.70$	& $53.00^{+0.26} _{-0.10}$	& $>0.20^{+0.07} _{-0.02}$\\ 
\gray	080319B	& $52.76$	& $555.70$	& $54.16\pm0.01$		& $>1.11\pm0.01$\\
	080514B	& $37.30$	& $174.70$	& $53.42^{+0.03} _{-0.04}$	& $>0.22\pm0.01$\\ 
\gray	080916C	& $97.49$	& $377.70$	& $54.49^{+0.08} _{-0.10}$	& $>0.80^{+0.08} _{-0.09}$\\ 
	090323	& $73.05$	& $1079.00$	& $54.61 \pm 0.01$		& $>1.68\pm0.02$\\ 
\midrule
\multicolumn{5}{l}{\textbf{\textbf{Unclear cases}}}\\
\midrule
\gray	050922C	& \dots 	& \dots		& $52.98^{+0.08} _{-0.05}$	& \dots\\ 
	051221A	& $25.92$	& $228.80$	& $51.41^{+0.02} _{-0.33}$	& \dots\\
\gray	060904B	& $3.58 $	& $162.80$	& $51.71^{+0.13} _{-0.21}$	& \dots\\ 
\bottomrule
\end{tabular}
\tablefoot{
Time intervals deduced from the lower panels in Fig.~\ref{fig:sample} if the first logarithmic
derivative of $F_{\rm opt}/F_{\rm x}$  points to either an ISM profile or a free wind, i.e., once the asymptotic value
$+1/4$ or $-1/4$ was reached (within the error bars). With the exception for GRB 080514B (\citealt {Rossi2009a})
and GRB 090726 (\citealt {Butler2010}), the isotropic equivalent energies, $E_{\rm iso}$, were taken from
\cite{Kann2008a, Kann2010a}. The forth column was calculated based on Eq. \ref{Radii}, assuming
$A_*=1$, with $t =\tilde{t}_{\rm start}$ (Fig. \ref{fig:sample}) for ISM profiles and $t = \tilde{t}_{\rm end}$
(Fig. \ref{fig:sample}) for wind profiles.}
\label{tab:radii}
\end{table}

%% file: 15581ampl.tex
\section{Smoothly broken power law of the order $m$}\label{app:mpl}

The equation for a smoothly broken power-law of the order $m$, $F_{\nu,m} (t)$,
was derived by recursion in the following way. Let us assume the function
$F_{\nu,m} (t)$ consists of $m$ power-law segments connected by $(m-1)$ breaks.
To add an additional power-law segment $\tilde F_{\nu,m+1} (t)$, we first
normalised the new power-law segment to the previous one, $F_{\nu, m}(t)$,
at the break time $t_{b,m}$:
\begin{equation}
\tilde{F}_{\nu, m+1}(t)=F_{\nu, m}(t_{b,m}) \left(\frac{t}{t_{b,m}}\right)^{-\alpha_{m+1}}\,.
\label{eq:1}
 \end{equation}
Here $\alpha_{m+1}$ is the slope of segment $(m+1)$. Second, we followed \cite{Beuermann1999a}
and introduced a smoothness parameter $n_m$ so that the smoothly broken power law of order the $(m+1)$ takes the form
\begin{equation}
F_{\nu, m+1}(t)=\left(F^{-n_{m}} _{\nu, m}(t) + \tilde{F}^{-n_{m}} _{\nu, m+1}(t)\right)^{-1/n_{m}}\,.
\label{eq:2}
 \end{equation}
If the light-curve consists of $m$ segments, both steps (adding and smoothing) have to
be performed $(m-1)$-times.

For example, let us derive the equation for a smoothly broken power-law \citep{Beuermann1999a}.
In this case $m=2$, thus the function consists of two power-law segments connected by one break
at the time $t_{b,1}$. The initial function is a simple power law $F_{\nu,1}(t)=C\,t^{-\alpha_1}$.
First, the second power-law segment, $\tilde{F}_{\nu,2}$, has to be connected to the first one at
the time $t_{b,1}$ (step \ref{eq:1})
\begin{eqnarray}
	\tilde{F} _{\nu,2} 	&=& F_{\nu,1}\left(t_{b,1}\right) \left(\frac{t}{t_{b,1}}\right)^{-\alpha_2}\nonumber\\
							&=& C t^{-\alpha_1} _{b,1} \left(\frac{t}{t_{b,1}}\right)^{-\alpha_2}\nonumber~.
\end{eqnarray}
Second, the transition has to be smoothed by weighting both functions at the point of intersection
(step \ref{eq:2})
\begin{eqnarray}
	F_{\nu,2} 	&=& \left( F^{-n_1} _{\nu,1}\left(t\right) + \tilde{F}^{-n_1} _{\nu,2} \left(t\right) \right)^{-1/n_1}\nonumber\\
				&=& \left( C^{-n_1} t^{\alpha_1\,n_1} + C^{-n_1} t^{\alpha_1\,n_1} _{b,1} \left(\frac{t}{t_{b,1}} \right)^{\alpha_2\,n_1} \right)^{-1/n_1}\nonumber\\
				&=& C\,t^{-\alpha_1} _{b,1} \left( \left(\frac{t}{t_{b,1}}\right)^{\alpha_1\,n_1} + \left(\frac{t}{t_{b,1}}\right)^{\alpha_2\,n_1}\right)^{-1/n_1}\nonumber ~.
\end{eqnarray}
This leads to the equation found by \cite{Beuermann1999a} for a smoothly broken power-law.
Repeating both steps leads to a smoothly broken power-law of the order 3 (double smoothly broken
power-law; \citealt{Liang2008a}). Thus, looping $(m-1)$-times over both steps results in a smoothly
broken power law of the order $m$.

%% file: 15581tbeta.tex
\begin{table*}
\caption{Properties of the afterglow SEDs in the optical and X-ray bands of the 27 bursts that entered our sample.}
\centering
\begin{tabular}{l c lccl ll}
\toprule																																					
\multirow{2}{*}{GRB}	&	\multirow{2}{*}{$z$}	&	\multicolumn{1}{c}{\multirow{2}*{$\beta_{\rm opt}$}}&	Dust	&	\multirow{2}{*}{$A^{\rm host} _{V}$}	&	\multicolumn{1}{c}{\multirow{2}*{$\beta_{\rm x}$}}&	\multicolumn{1}{c}{$t_{\rm mid}$}	&	\multirow{2}{*}{References}	\\
						&							&									&	Model	&											&									&	\multicolumn{1}{c}{(ks)}			&								\\
\midrule																																					

\gray	050603		&$2.818	$&$	\dots		$&\dots		&$	\dots			$&$0.96	\pm	0.12	$&	X: 44		&$z$: 1; Opt: 2		\\
	050801		&$1.560	$&$0.69	\pm	0.34	$&	SMC	&$0.30	\pm	0.18		$&$0.89	\pm	0.08	$&	Opt: 86; X: 79	&$z$: 3; Opt: 2		\\
\gray	050820A		&$2.615	$&$0.72	\pm	0.03	$&	SMC	&$0.07	\pm	0.01		$&$1.09	\pm	0.03	$&	Opt: 86; X: 37	&$z$: 4; Opt: 2		\\
	050922C$\dag$	&$2.199	$&$1.07			$&	MW	&$0.17	\pm	0.05		$&$1.07	\pm	0.05	$&	Opt/X: 20	&$z$: 6; Opt/X: 5	\\
\gray	051109A$\dag$	&$2.346	$&$0.40			$&	SMC	&$<0.10				$&$0.90	\pm	0.04	$&	Opt/X: 5	&$z$: 7; Opt/X: 5	\\

	051221A		&$0.546	$&$	\dots		$&\dots		&$	\dots			$&$0.95	\pm	0.11	$&	X: 3		&$z$: 8; Opt: 9		\\
\gray	060418		&$1.490	$&$0.69	\pm	0.11	$&	LMC	&$0.20	\pm	0.08		$&$0.98	\pm	0.15	$&	Opt: 86; X: 7	&$z$: 10; Opt: 2	\\
	060512		&$0.443	$&$	\dots		$&\dots		&$	\dots			$&$1.02	\pm	0.11	$&	X: 23		&$z$: 11		\\
\gray	060614		&$0.125	$&$0.81 \pm 0.08	$&	SMC	&$0.05 \pm 0.02 		$&$0.81	\pm	0.11	$& 	Opt: 67; X: 36	&$z$: 12; Opt: 13	\\
	060714$\dag$	&$2.711	$&$0.92			$&	SMC	&$0.46	\pm	0.17		$&$0.92	\pm	0.10	$&	Opt/X: 5	&$z$: 14; Opt/X: 5	\\

\gray	060904B		&$0.703	$&$1.11	\pm	0.10	$&	SMC	&$0.08	\pm	0.08		$&$1.11	\pm	0.12	$&	Opt: 86; X: 2 	&$z$: 4; Opt: 2		\\
	060908		&$1.884	$&$0.30	\pm	0.03	$&	SMC	&$0.00				$&$0.94	\pm	0.08	$&	Opt: 86; X: 2	&$z$: 4; Opt: 2		\\
\gray	070411		&$2.954	$&$	\dots		$&		&$	\dots			$&$1.19	\pm	0.14	$&	X: 8		&$z$: 4				\\
	070802$\dag$	&$2.454	$&$0.61			$&	MW	&$1.20	\pm	0.12		$&$1.11	\pm	0.05	$&	Opt/X: 2	&$z$: 4; Opt/X: 15	\\

\gray	070810A		&$2.170	$&$	\dots		$&\dots		&$	\dots			$&$1.14	\pm	0.16	$&	X:	2	&$z$: 16			\\
	080319B		&$0.937	$&$0.50		\pm 0.07$&	SMC	&$ 0.15				$&$0.98	\pm 0.10	$& 	Opt/X: 170	&$z$: 17; Opt/X: 18 	\\
\gray	080514B$\dag$	&$1.800	$&$0.63	\pm	0.02	$&	SMC	&$0.00				$&$1.13	\pm 0.13	$&	Opt: 43; X: 33	&$z$/Opt/X:	19		\\
	080710$\dag$	&$0.845	$&$1.01			$&	SMC	&$0.00	\pm	0.00		$&$1.01	\pm	0.01	$&	Opt/X: 27	&$z$: 4; Opt/X: 20	\\
\gray	080721		&$2.591	$&$	\dots		$&\dots		&$	\dots			$&$0.99	\pm	0.04	$&	X: 4		&$z$: 4				\\

	080916C$\dag$	&$4.350	$&$0.49			$&	SMC	&$0.00				$&$0.49	\pm	0.34	$&	Opt: 79; X: 67	&$z$/Opt/X: 21		\\
\gray	081203A		&$2.050	$&$	\dots		$&\dots		&$	\dots			$&$1.06	\pm	0.07	$&	X: 22		&$z$: 22			\\
	090102		&$1.547	$&$0.74	\pm	0.22	$&	SMC	&$0.12	\pm	0.11		$&$0.77	\pm	0.04	$&	Opt: 86; X: 8	&$z$: 23; Opt: 2	\\
\gray	090323		&$3.568	$&$0.65	\pm	0.13	$&	SMC	&$0.14	\pm	0.04		$&$0.95	\pm	0.13	$&	Opt: 97 ks, X: 92 ks&$z$: 24; Opt: 25\\
	090328		&$0.735	$&$1.17	\pm	0.17	$&	SMC	&$0.18	\pm	0.13		$&$1.10	\pm	0.16	$&	Opt: 86; X: 130	&$z$: 25; Opt: 2	\\

\gray	090726		&$2.710	$&$	\dots		$&\dots	&$	\dots				$&$1.45	\pm	0.15	$&	X: 17		&$z$: 26			\\
	090902B		&$1.822	$&$0.73	\pm	0.13	$&	SMC	&$0.05	\pm	0.05		$&$1.02	\pm	0.11	$&	Opt: 86; X: 98	&$z$: 24; Opt: 2	\\
\gray	090926A$\dag$	&$2.106	$&$1.04			$&	MW	&$<0.10				$&$1.04	\pm	0.08	$&	Opt/X: 250	&$z$:24; Opt/X:	27		\\
\bottomrule
\end{tabular}
 \tablefoot{For every GRB we list the redshift, $z$, the properties of the optical SED
 	(spectral slope $\beta_{\rm opt}$, host extinction in the $V$-band rest frame, $A^{\rm host} _{\rm V}$,
 	and the corresponding extinction law), the spectral slope in the X-rays, $\beta_{\rm x}$,
 	and the mean time after the burst, $t_{\rm mid}$, at which the SED was extracted.
	The events for which we used joint optical-to-X-ray SEDs are marked with a '$\dag$'.
	In these cases the difference in the spectral slope, $\beta_{\rm x} - \beta_{\rm opt}$ was
	either fixed to $1/2$ or 0. Except for GRB 080514B, these estimates have only
	one error estimate for $\beta_{\rm opt}$ and $\beta_{\rm x}$ together due to the fitting
	procedure described in the associated papers.
}
\tablebib{	
(1)\,	\citet{Berger2005a};
(2)\,	\citet{Kann2010a};
(3)\,	\citet{Pasquale2007a};
(4)\,	\citet{Fynbo2009a};
(5)\,	\citet{Schady2010a};
(6)\,	\citet{Piranomonte2008a};
(7)\,	\citet{Quimby2005a};
(8)\,	\citet{Soderberg2006a};
(9)\,	\citet{Kann2008a};
(10)\,	\citet{Prochaska2006a};
(11)\,	\citet{Bloom2006a};
(12)\,	\citet{DellaValle2006b};
(13)\,	\citet{Mangano2007a};
(14)\,	\citet{Jakobsson2006a};
(15)\,	\citet{Kruehler2008a};
(16)\,	\citet{Thoene2007a};
(17)\,	\citet{DElia2009a};
(18)\,	\citet{Racusin2008a};
(19)\,	\citet{Rossi2009a};
(20)\,	\citet{Kruehler2009a};
(21)\,	\citet{Greiner2009b};
(22)\,	\citet{Kuin2009a};
(23)\,	\citet{Postigo2009a};
(24)\,	\citet{Cenko2010a};
(25)\,	\citet{McBreen2010a};
(26)\,	\citet{Fatkhullin2009a};
(27)\,	\citet{Rau2010a}
}
\label{tab:sed}
\end{table*}

%% file: 15581talfa.tex
\begin{table*}
\caption{Light-curve parameters of the late-time optical and X-ray afterglows of the 27 bursts that entered our sample.}
\centering
\begin{tabular}{l c rr c c r rr}
\toprule
\multirow{2}{*}{GRB}								&	\multirow{2}{*}{Band}	& \multicolumn{5}{c}{Light-curve parameters}																				&\multicolumn{2}{c}{Overlapping time interval}			\\
													&							& $t^{\rm late} _{\rm start}$ (ks)	& $t^{\rm late} _{\rm end}$ (ks)	&	$\alpha_1$			&	$\alpha_2$			&\multicolumn{1}{c}{$t_{\rm b, jet}$ (ks)}	&$t_{\rm start}$	(ks)	&	$t _{\rm end}$ (ks)	\\
\midrule

\gray							&	O	&	33.33	&	196.67	&$		\dots		$&$1.97\pm	0.06$	&							&							&									\\
\gray\multicolumn{1}{g}{\multirow{-2}*{050603	}}	&	X	&	39.21	&	577.04	&$		\dots		$&$1.67\pm	0.05$	&\multirow{-2}*{$<	39.2$}	&\multirow{-2}*{39.2}	& \multirow{-2}*{196.7}				\\

							&	O	&	0.25	&	106.27	&$	1.19	\pm	0.01	$&$	\dots		$&							&							&									\\
\multirow{-2}*{050801	}				&	X	&	0.22	&	448.10	&$	1.29	\pm	0.08	$&$	\dots		$&\multirow{-2}*{$>	106.3$}	&\multirow{-2}*{0.25}		&\multirow{-2}*{106.3}				\\

\gray							&	O	&	23.89	&	612.10	&$	1.04	\pm	0.01	$&$	\dots		$&$\dots$					&							&									\\
\gray\multicolumn{1}{g}{\multirow{-2}*{050820A	}}	&	X	&	4.64	&	3676.50	&$	1.19	\pm	0.02	$&$1.73\pm	0.16	$&$	640.2	\pm	138.7	$	&\multirow{-2}*{23.9}	&\multirow{-2}*{612.1}			\\

							&	O	&	4.13	&	606.01	&$	1.47	\pm	0.04	$&$	\dots		$&							&							&									\\
\multirow{-2}*{050922C	}				&	X	&	2.02	&	95.92	&$	1.37	\pm	0.03	$&$	\dots		$&\multirow{-2}*{$>95.9$}	&\multirow{-2}*{4.1}		&\multirow{-2}*{95.9}				\\

\gray							&	O	&	17.52	&	265.20	&$	1.03	\pm	0.06	$&$	\dots		$&							&							&									\\
\gray\multicolumn{1}{g}{\multirow{-2}*{051109A	}}	&	X	&	3.46	&	1240.80	&$	1.20	\pm	0.01	$&$	\dots		$&\multirow{-2}*{$>265.2$}	&\multirow{-2}*{17.5}		&\multirow{-2}*{265.2}				\\

							&	O	&	4.68	&	445.12	&$	0.96	\pm	0.03	$&$	\dots		$&$\dots$					&							&									\\
\multirow{-2}*{051221A	}				&	X	&	15.17	&	2271.92	&$	1.03	\pm	0.01	$&$1.99\pm	0.17	$&$	318.5	\pm	38.6	$	&\multirow{-2}*{15.2}		&\multirow{-2}*{445.1}				\\

\gray							&	O	&	0.12	&	871.05	&$	1.192	\pm	0.002	$&$	\dots		$&							&							&									\\
\gray\multicolumn{1}{g}{\multirow{-2}*{060418	}}	&	X	&	4.40	&	537.08	&$	1.52	\pm	0.05	$&$	\dots		$&\multirow{-2}*{$>537.1$}	&\multirow{-2}*{4.4}		& \multirow{-2}*{537.1}				\\

							&	O	&	7.02	&	715.56	&$	0.80	\pm	0.03	$&$	\dots		$&							&							&									\\
\multirow{-2}*{060512	}				&	X	&	0.24	&	288.61	&$	1.12	\pm	0.05	$&$	\dots		$&\multirow{-2}*{$>288.6$}	&\multirow{-2}*{7.0}		&\multirow{-2}*{288.6}				\\

\gray							&	O	&	51.59	&	1700.19	&$	1.05	\pm	0.04	$&$2.45\pm	0.05	$&$	113.1	\pm	2.7		$	&							&									\\
\gray\multicolumn{1}{g}{\multirow{-2}*{060614	}}	&	X	&	41.23	&	1500.57	&$	1.07	\pm	0.11	$&$2.31\pm	0.11	$&$	120.0	\pm	12.1	$	&\multirow{-2}*{51.6}		&\multirow{-2}*{1500.6}				\\

							&	O	&	47.07	&	286.77	&$	1.42	\pm	0.18	$&$	\dots		$&$						$	&							&	\\
\multirow{-2}*{060714	}				&	X	&	3.07	&	1029.03	&$	1.23	\pm	0.03	$&$	\dots		$&\multirow{-2}*{$>286.8$}	&\multirow{-2}*{47.1}		&\multirow{-2}*{286.8}				\\

\gray							&	O	&	3.52	&	163.13	&$	1.16	\pm	0.02	$&$	\dots		$&							&							&									\\
\gray\multicolumn{1}{g}{\multirow{-2}*{060904B	}}	&	X	&	3.58	&	391.01	&$	1.36	\pm	0.03	$&$	\dots		$&\multirow{-2}*{$>163.1$}	&\multirow{-2}*{3.6}		&\multirow{-2}*{163.1}				\\

							&	O	&	0.14	&	82.42	&$	1.03	\pm	0.01	$&$	\dots		$&							&							&									\\
\multirow{-2}*{060908	}				&	X	&	0.71	&	483.21	&$	1.49	\pm	0.07	$&$	\dots		$&\multirow{-2}*{$>82.4$}	&\multirow{-2}*{0.71}		& \multirow{-2}*{82.4}				\\

\gray							&	O	&	92.09	&	523.12	&$	1.29	\pm	0.06	$&$	\dots		$&							&							&									\\
\gray\multicolumn{1}{g}{\multirow{-2}*{070411	}}				&	X	&	0.50	&	664.68	&$	1.12	\pm	0.02	$&$	\dots		$&\multirow{-2}*{$>523.1$}	&\multirow{-2}*{92.1}		& \multirow{-2}*{523.1}				\\

								&	O	&	10.90	&	88.88	&$	0.90	\pm	0.16	$&$	\dots		$&							&							&									\\
\multirow{-2}*{070802	}	&	X	&	6.68	&	316.52	&$	1.17	\pm	0.09	$&$	\dots		$&\multirow{-2}*{$>88.9$}	&\multirow{-2}*{10.9}		&\multirow{-2}*{88.9}				\\

\gray							&	O	&	1.95	&	103.36	&$	1.30	\pm	0.10	$&$	\dots		$&							&							&									\\
\gray\multicolumn{1}{g}{\multirow{-2}*{070810A	}}				&	X	&	1.55	&	34.46	&$	1.29	\pm	0.07	$&$	\dots		$&\multirow{-2}*{$>34.5$}	&\multirow{-2}*{1.9}		&\multirow{-2}*{34.5}				\\
							&	O	&	0.8	&	1060	&$ 1.237	\pm 0.002	$&$	 \dots		$&							&							&									\\
\multirow{-2}*{080319B	}	&	X	&	42.66	&	2559.17	&$ 1.04		\pm 0.05	$&$2.68\pm 0.36		$&$953\pm 137			$	&\multirow{-2}*{42.7}			&\multirow{-2}*{1060}				\\
\bottomrule
\end{tabular}
\tablefoot{
	Columns 3 to 7 give the time interval, $t^{\rm late} _{\rm start}, t^{\rm late} _{\rm stop}$, the pre- and post-jet break decay slopes,
	$\alpha_1$, $\alpha_2$, and the observed jet break time, $t_{b, \rm jet}$. 
	The optical and X-ray light curves were independently fitted for every GRB (Sect. \ref{sec:sample}).
	Owing this, two break times are given for GRB 060614. Within the errors the break was achromatic,
	making it a good candidate for a jet break. For the other GRBs we state the upper or lower limits on the jet
	break time with respect to the identification of the dynamical regime shown in 	Table \ref{tab:cbm}.
	The overlapping time interval of the late-time optical and X-ray afterglow is shown in the last two columns.
	$\dag$ The optical afterglow light curve of GRB 080721 shows an additional shallow break at ($129\pm84$) ks. The difference
	in the pre- and post-break decay slope is $\approx0.25$ in agreement with \citet{Kann2010a}, which is typical for a cooling break.
	This break is not a jet break.
	}
\label{tab:input}
\end{table*}

\begin{table*}
\centering
\begin{tabular}{l c rr c c r rr}
\toprule
\multirow{2}{*}{GRB}								&	\multirow{2}{*}{Band}	& \multicolumn{5}{c}{Light-curve parameters}																				&\multicolumn{2}{c}{Overlapping time interval}			\\
													&							& $t_{\rm start}$ (ks)	& $t _{\rm end}$ (ks)	&	$\alpha_1$			&	$\alpha_2$			&\multicolumn{1}{c}{$t_{\rm b, jet}$ (ks)}	&$t_{\rm start}$	(ks)	&	$t _{\rm end}$ (ks)	\\
\midrule

\gray							&	O	&	37.15	&	174.79	&$	1.64	\pm	0.06	$&$	\dots		$&							&							&									\\
\gray\multicolumn{1}{g}{\multirow{-2}*{080514B	}}				&	X	&	37.30	&	217.27	&$	1.54	\pm	0.14	$&$	\dots		$&\multirow{-2}*{$>174.8$}	&\multirow{-2}*{37.3}		&\multirow{-2}*{174.8}				\\

							&	O	&	9.97	&	353.11	&$	1.57	\pm	0.02	$&$	\dots		$&							&							&									\\
\multirow{-2}*{080710	}	&	X	&	11.34	&	349.31	&$	1.56	\pm	0.09	$&$	\dots		$&\multirow{-2}*{$>349.3$}	&\multirow{-2}*{11.3}		&\multirow{-2}*{349.3}				\\
\gray							&	O	&	0.17	&	2641.08	&$	1.22	\pm	0.01	$&$	1.46	\pm	0.08		$&							&							&									\\
\gray\multicolumn{1}{g}{\multirow{-2}*{080721$\dag$	}}				&	X	&	29.46	&	1256.83	&$	1.50	\pm	0.03	$&$	\dots		$&\multirow{-2}*{$>1256.8$}	&\multirow{-2}*{29.5}		&\multirow{-2}*{1256.8}				\\

							&	O	&	97.49	&	377.94	&$	1.40	\pm	0.10	$&$	\dots		$&							&								&								\\
\multirow{-2}*{080916C	}	&	X	&	65.03	&	1306.33	&$	1.29	\pm	0.08	$&$	\dots		$&\multirow{-2}*{$>377.9$}	&\multirow{-2}*{97.5}		&\multirow{-2}*{377.9}				\\

\gray							&	O	&	6.60	&	301.54	&$	\dots			$&$1.72\pm	0.01	$&$\dots$					&							&									\\
\gray\multicolumn{1}{g}{\multirow{-2}*{081203A	}}				&	X	&	8.91	&	345.00	&$	1.13	\pm	0.01	$&$1.93\pm	0.06	$&$	8.9	\pm	0.9			$	&\multirow{-2}*{8.9}		&\multirow{-2}*{301.5}				\\

							&	O	&	15.76	&	263.68	&$	1.50	\pm	0.03	$&$	\dots		$&							&							&									\\
\multirow{-2}*{090102	}	&	X	&	0.95	&	688.48	&$	1.43	\pm	0.02	$&$	\dots		$&\multirow{-2}*{$>263.6$}	&\multirow{-2}*{15.8}		&\multirow{-2}*{263.7}				\\

\gray							&	O	&	96.81	&	1142.30	&$	\dots			$&$	1.88\pm	0.01	$&							&							&									\\
\gray\multicolumn{1}{g}{\multirow{-2}*{090323	}}				&	X	&	70.47	&	1084.77	&$	\dots			$&$	1.56\pm	0.10	$&\multirow{-2}*{$<96.8$}	&\multirow{-2}*{96.8}		&\multirow{-2}*{1084.8}				\\

							&	O	&	57.28	&	1007.14	&$	\dots			$&$1.78	\pm	0.04$	&							&							&									\\
\multirow{-2}*{090328	}	&	X	&	0.15	&	924.60	&$	\dots			$&$1.68	\pm	0.09$	&\multirow{-2}*{$<57.3$}	&\multirow{-2}*{57.3}		& \multirow{-2}*{924.6}				\\

\gray							&	O	&	1.40	&	10.00	&$	0.97	\pm	0.10	$&$	\dots		$&							&							&									\\
\gray\multicolumn{1}{g}{\multirow{-2}*{090726	}}				&	X	&	3.61	&	66.43	&$	1.34	\pm	0.04	$&$	\dots		$&\multirow{-2}*{$>10.0$}	&\multirow{-2}*{3.6}		&\multirow{-2}*{10.0}				\\

							&	O	&	45.64	&	1171.46	&$	0.97	\pm	0.02	$&$	\dots		$&$				$			&							&									\\
\multirow{-2}*{090902B	}	&	X	&	45.21	&	1456.69	&$	1.33	\pm	0.03	$&$	\dots		$&\multirow{-2}*{$>1171.5$}	&\multirow{-2}*{45.6}		& \multirow{-2}*{1171.5}			\\

\gray							&	O	&	254.24	&	2070.84	&$	1.74	\pm	0.02	$&$	\dots		$&							&							&									\\
\gray\multicolumn{1}{g}{\multirow{-2}*{090926A	}}				&	X	&	51.49	&	1803.53	&$	1.53	\pm	0.07	$&$	\dots		$&\multirow{-2}*{$>1803.5$}	&\multirow{-2}*{254.2}		&\multirow{-2}*{1803.5}				\\

\bottomrule
\end{tabular}
\hfill\center{Table \ref{tab:input} ---  continued}
\end{table*}

%% file: 15581tres.tex
\begin{table*}
\caption{Identification of the light-curve segments and the circumburst medium.}
\centering
\begin{tabular}{l l rr  lll l|l ll}
\toprule
\multirow{2}{*}{GRB}	& \multicolumn{1}{c}{\multirow{2}*{Closure relation}}	&	\multicolumn{1}{c}{\multirow{2}*{$\alpha_{\rm x} - \alpha_{\rm opt}$}}	& \multicolumn{1}{c}{\multirow{2}*{$\beta_{\rm x} - \beta_{\rm opt}$}}	&&	\multicolumn{2}{c}{\multirow{2}*{Conclusion}}	&& \multicolumn{2}{c}{Literature}\\
						&														&																			&																		&&													&&& ID	& Reference\\
\midrule

\gray						&O: ---				&  						& 					&& O: j1b	& 				&&&\\
\gray\multicolumn{1}{g}{\multirow{-2}*{050603}}	&X: j1/2a, j2b			& \multirow{-2}*{$-0.30 \pm 0.08$}		& \multirow{-2}{*}{---}			&& X: j2b	& \multirow{-2}{*}{Wind}	&&&\\

\multirow{2}{*}{050801}				&O: 				&  						& 					&& O: S1a	& 				&& O: S1a	& \\
						&X: S1a				& \multirow{-2}{*}{$0.10 \pm 0.08$} 		& \multirow{-2}{*}{$0.20\pm 0.35$}	&& X: S1a	& \multirow{-2}*{ISM}		&& X: S1a	& \multirow{-2}{*}{1}\\

\gray						&O: S1a				&  						& 					&& O: S1a	& 				&& O: S1a	& \\
\gray						&X$_1$: S2 			& \multirow{-2}{*}{$0.15 \pm 0.02$}		& \multirow{-2}*{$0.37\pm 0.04$}	&& X$_1$: S2	& 				&& X$_1$: S2& \\
\gray\multicolumn{1}{g}{\multirow{-3}*{050820A}}&X$_2$: J2, j2a/b	 	& 						&					&& X$_2$: j2a	& \multirow{-3}{*}{ISM}		&& X$_2$: ---& \multirow{-3}{*}{2}\\

\multirow{2}{*}{050922C}			&O: S1a				& 						& 					&& O: S1a	& 				&&&\\
						&X: S1a				& \multirow{-2}{*}{$-0.10 \pm 0.05$}		& \multirow{-2}{*}{$0.00 \pm 0.05$}	&& X: S1a	& \multirow{-2}{*}{ISM}		&&&\\

\gray						&O: 				&  						& 					&& O: S1a	& 				&&&\\	
\gray\multicolumn{1}{g}{\multirow{-2}*{051109A}}&X: S1a, S2			& \multirow{-2}{*}{$0.17 \pm 0.06$}		& \multirow{-2}{*}{$0.50 \pm 0.04$}	&& X: S2	& \multirow{-2}*{ISM}		&&& \\

						&O: ---				& 						& 					&& O: S2	& 				&&\\
						&X$_1$: S1a, S2			& \multirow{-2}{*}{$0.07 \pm 0.03$} 		& \multirow{-2}{*}{	---}		&& X$_1$: S2	& 				&&X$_1$: S2\\
\multirow{-3}*{051221A}				&X$_2$: J2, j1/2a, j1/2b	& 						&					&& X$_2$: J2	& \multirow{-3}{*}{---}		&&& \multirow{-3}{*}{3}\\

\gray						&O: S1a				&  						& 					&& O: S1a	& 				&&& \\	
\gray\multicolumn{1}{g}{\multirow{-2}*{060418}}	&X: S1a/b, S2			& \multirow{-2}{*}{$0.33 \pm 0.05$}		& \multirow{-2}{*}{$0.29 \pm 0.19$}	&& X: S2	& \multirow{-2}*{ISM}		&&&  \\

\multirow{2}{*}{060512}				&O: ---				& 						& 					&& O: S1a	& 				&&&\\
						&X: S1a, S2			& \multirow{-2}{*}{$0.32 \pm 0.06$} 		& \multirow{-2}{*}{---}			&& X: S2	& \multirow{-2}{*}{ISM}		&&&\\

\gray						&O$_1$: S1a, S2			&  						& 					&& O$_1$: S1a 	& 				&& &\\
\gray						&X$_1$: S1a, S2			& \multirow{-2}{*}{$0.02 \pm 0.12$}		& \multirow{-2}{*}{$0.00 \pm 0.11$}	&& X$_1$: S1a 	& 				&& X$_1$: S1a&\\
\gray						&O$_2$: J1, j1b			&  						& 					&& O$_2$: J1 	& 				&& &\\
\gray\multicolumn{1}{g}{\multirow{-4}*{060614}}	&X$_2$: J1/2, j1a/b		& \multirow{-2}{*}{$-0.14 \pm 0.12$}		& \multirow{-2}{*}{$0.00 \pm 0.11$}	&& X$_2$: J1	& \multirow{-4}*{ISM}		&& X$_2$: J1	& \multirow{-4}{*}{4}\\

						&O: S1a/b, S2			& 						& 					&& O: S1a	& 				&&&\\
\multirow{-2}{*}{060714}			&X: S1a, S2			& \multirow{-2}{*}{$-0.19 \pm 0.18$}		& \multirow{-2}{*}{$0.00 \pm 0.10$}	&& X: S1a	& \multirow{-2}*{ISM}		&&&\\

\gray						&O: S2				& 						& 					&& O: S2	& 				&&&\\
\gray\multicolumn{1}{g}{\multirow{-2}*{060904B}}&X: S1a, S2			& \multirow{-2}{*}{$0.20 \pm 0.04$} 		& \multirow{-2}{*}{$0.00 \pm 0.16$}	&& X: S2	& \multirow{-2}*{---}		&&&\\

						&O: ---				&  						& 					&& O: S1a	& 				&&& \\		
\multirow{-2}{*}{060908}			&X: S1a				& \multirow{-2}{*}{$0.46 \pm 0.07$}		& \multirow{-2}*{$0.64 \pm 0.09$}	&& X: S2	& \multirow{-2}*{ISM}		&&& \\

\gray						&O: ---				&  						& 					&& O: S1b	& 				&&&\\
\gray\multicolumn{1}{g}{\multirow{-2}*{070411}}	&X: S2				& \multirow{-2}{*}{$-0.17 \pm 0.06$}		& \multirow{-2}*{---}			&& X: S2	& \multirow{-2}*{Wind}		&&&\\

\bottomrule
\end{tabular}
\tablefoot{Following the scheme in Sect. \ref{sec:ensemble}, different criteria were applied to find a
	consistent description of the late-time optical and X-ray afterglow data to reveal the nature
	of the circumburst medium. The light-curve segments were labelled following \citet[][for the designated ID see also Table \ref{tab:closure}]{Panaitescu2007a}
	((S, J, j)=(spherical expansion, jet with sideways expansion, jet without lateral spreading),
	  (1, 2)=($\nu<\nu_c$, $\nu>\nu_c$), (a, b)=(ISM, wind); see also Table \ref{tab:closure}). If there was a break in the light curve,
	the segments were tagged with the subscript 1 and 2.
	The second column lists the light-curve segments and all closure relations that agree
	with the observational data within $3\sigma$. The third and fourth column show the
	corresponding measured differences in the temporal and spectral slopes, if available. Their uncertainties are $1\sigma$ errors.
	If a criterion could not be applied or a solution was not found, we crossed the field out.
	The fifth and sixth columns show the conclusion based on all criteria, i.e., the segment
	label and the type of circumburst medium. The second last column summarises the results from the literature.
}
\tablebib{
 (1)\, \citet{Rykoff2006a};
 (2)\, \citet{Cenko2006a};
 (3)\, \citet{Burrows2006a};
 (4)\, \citet{Mangano2007a}
}
\label{tab:cbm}
\end{table*}

\begin{table*}
\centering
\begin{tabular}{l l rr  lll l|l ll}
\toprule
\multirow{2}{*}{GRB}	& \multicolumn{1}{c}{\multirow{2}*{Closure relation}}	&	\multicolumn{1}{c}{\multirow{2}*{$\alpha_{\rm x} - \alpha_{\rm opt}$}}	& \multicolumn{1}{c}{\multirow{2}*{$\beta_{\rm x} - \beta_{\rm opt}$}}	&&	\multicolumn{2}{c}{\multirow{2}*{Conclusion}}	&& \multicolumn{2}{c}{Literature}\\
						&														&																			&																		&&													&&& ID	& Reference\\
\midrule

						&O: S1a/b, S2			& 						& 					&& O: S1a	& 				&& O: S2	&\\
\multirow{-2}{*}{070802}			&X: S2				& \multirow{-2}{*}{$0.27 \pm 0.18$} 		& \multirow{-2}{*}{$0.50 \pm 0.05$}	&& X: S2	& \multirow{-2}{*}{ISM} 	&& X: S2 	&\multirow{-2}{*}{5}\\

\gray						&O: ---				&  						& 					&& O: S1a	& 				&&&\\	
\gray\multicolumn{1}{g}{\multirow{-2}*{070810A}}&X: S1a, S2			& \multirow{-2}{*}{$-0.01 \pm 0.12$}		& \multirow{-2}*{---}			&& X: S2	& \multirow{-2}*{ISM}		&&&\\

						&O: S1b				&						&					&& O: S1b	& 				&& O: S1b	&\\
						&X$_1$:S1a, S2			& \multirow{-2}{*}{$-0.20 \pm 0.05$}		& \multirow{-2}*{$0.48 \pm 0.12$}	&& X$_1$: S2	& 				&& X$_1$: S2&\\
\multirow{-3}{*}{080319B}			&X$_2$:				&						&					&& X$_2$: ---	&\multirow{-3}*{Wind}		&& X$_2$: J2&\multirow{-3}{*}{6}\\

\gray						&O: S1b				&  						& 					&& O: S1b	& 				&& O: S1b	&\\
\gray\multicolumn{1}{g}{\multirow{-2}*{080514B}}&X: S1a/b, S2			& \multirow{-2}{*}{$-0.10 \pm 0.14$}		& \multirow{-2}{*}{$0.50 \pm 0.13$}	&& X: S2	& \multirow{-2}{*}{Wind}	&& X: S2	&\multirow{-2}{*}{7}\\

						&O: S1a, j2b			&  						& 					&& O: S1a	& 				&& O: S1a	&\\
\multirow{-2}{*}{080710}			&X: S1a, j2a/b			& \multirow{-2}{*}{$-0.01 \pm 0.09$}		& \multirow{-2}{*}{$0.00 \pm 0.01$}	&& X: S1a	& \multirow{-2}{*}{ISM}		&& X: S1a	&\multirow{-2}{*}{8}\\

\gray						&O$_1$: ---			&  						& 					&& O$_1$: S1a	&				&& 1) O$_1$: S1a; 2) O: S1a	& \\
\gray						&O$_2$: ---			&  						& 					&& O$_2$: S2	&				&& 1) O$_2$: S2			& \\
\gray\multicolumn{1}{g}{\multirow{-3}*{080721}}	&X: S2				& \multirow{-2}{*}{$0.04 \pm 0.09$}		& \multirow{-2}{*}{---}			&& X: S2	& \multirow{-3}{*}{ISM}		&& 1) X: ---; 2) X: S1a		& \multirow{-3}{*}{1) 9; 2) 10}\\

						&O: 				&  						& 					&& O: S1b	& 				&& O: S1a/b	&\\
\multirow{-2}{*}{080916C}			&X: 				& \multirow{-2}{*}{$-0.11 \pm 0.13$}		& \multirow{-2}*{$0.00\pm 0.34$}	&& X: S1b	& \multirow{-2}{*}{Wind}	&& X: S1a/b	& \multirow{-2}{*}{11}\\

\gray						&X$_1$: S2			&  						&					&& X$_1$: S2	& 				&&&\\
\gray						&O: ---				& 						& \multirow{-2}{*}{---}			&& O: j1a	& 				&&&\\
\gray\multicolumn{1}{g}{\multirow{-3}*{081203A}}&X$_2$: J2, j2a/b		& \multirow{-2}{*}{$0.21 \pm 0.06$}		&					&& X$_2$: j2a	& \multirow{-3}{*}{ISM}		&&&\\

						&O: 				&  						& 					&& O: S1a	& 			&& O: S1a		& \\
\multicolumn{1}{l}{\multirow{-2}*{090102}}	&X: 				& \multirow{-2}{*}{$0.07 \pm 0.04$}		& \multirow{-2}*{$0.02 \pm 0.22$}	&& X: S1a	& \multirow{-2}{*}{ISM}	&& X: S1a		& \multirow{-2}{*}{12}\\

\gray						&O: S1b, j1a/b			&  						& 					&& O: j1b	& 			&& 1) O: J1, j1a/b; 2) O: S1b &\\
\gray\multicolumn{1}{g}{\multirow{-2}*{090323}}	&X: S1a/b, S2, j1/2a, j2b	& \multirow{-2}{*}{$-0.32 \pm 0.10$}		& \multirow{-2}{*}{$0.30 \pm 0.18$}	&& X: j2b	& \multirow{-2}{*}{Wind}&& 1) X: ---; 2) X: S1b	 & \multirow{-2}{*}{1) 13; 2) 14}\\

						&O: S1a/b, S2, j1a, j2a/b	&  						& 					&& O: j2a	& 			&& O: J1		&\\
\multicolumn{1}{l}{\multirow{-2}*{090328}}	&X: S1a/b, S2, j1a, j2a/b	& \multirow{-2}{*}{$-0.10 \pm 0.09$}		& \multirow{-2}{*}{$-0.07 \pm 0.21$}	&& X: j2a	& \multirow{-2}{*}{ISM}	&& X: J1		&\multirow{-2}{*}{13}\\

\gray						&O: ---				&  						& 					&& O: S1a	& 			&& O: S1a		&\\
\gray\multicolumn{1}{g}{\multirow{-2}*{090726}}	&X: S2				& \multirow{-2}{*}{$0.37 \pm 0.10$}		& \multirow{-2}{*}{---}			&& X: S2	& \multirow{-2}{*}{ISM}	&& X: S2		&\multirow{-2}{*}{15}\\

						&O: S1a, S2			& 						& 					&& O: S1a	& 			&& O: S1a		&\\
  \multicolumn{1}{l}{\multirow{-2}*{090902B}}	&X: S1a, S2			& \multirow{-2}{*}{$0.36 \pm 0.04$}		& \multirow{-2}*{$0.29 \pm 0.17$}	&& X: S2	& \multirow{-2}{*}{ISM}	&& X: S2		&\multirow{-2}{*}{13, 14}\\
	
\gray						&O: S1a/b			& 						& 					&& O: S1a	& 			&& 1) O: S2; 2) O: S1a	&\\
\gray\multicolumn{1}{g}{\multirow{-2}*{090926A}}&X: S1a				& \multirow{-2}{*}{$-0.21 \pm 0.07$}		& \multirow{-2}{*}{$0.00 \pm 0.08$}	&& X: S1a	& \multirow{-2}{*}{ISM}	&& 1) X: S2; 2) X: S1a	& \multirow{-2}{*}{1) 14: 2) 16}\\

\bottomrule
\end{tabular}
\tablebib{
 (5)\, \citet{Kruehler2008a};
 (6)\, \citet{Racusin2008a};
 (7)\, \citet{Rossi2009a};
 (8)\, \citet{Kruehler2009a};
 (9)\, \citet{Kann2010a};
 (10)\, \citet{Starling2009a};
 (11)\, \citet{Greiner2009a};
 (12)\, \citet{Gendre2010a};
 (13)\, \citet{McBreen2010a};
 (14)\, \citet{Cenko2010a};
 (15)\, \citet{Simon2010a};
 (16)\, \citet{Rau2010a}
 }
\hfill\center{Table \ref{tab:cbm} ---  continued}
\end{table*}

%% file: 15581fig.tex
\begin{figure*}
\includegraphics[bb=44 217 654 835, clip, width=0.5\textwidth, angle=-90]{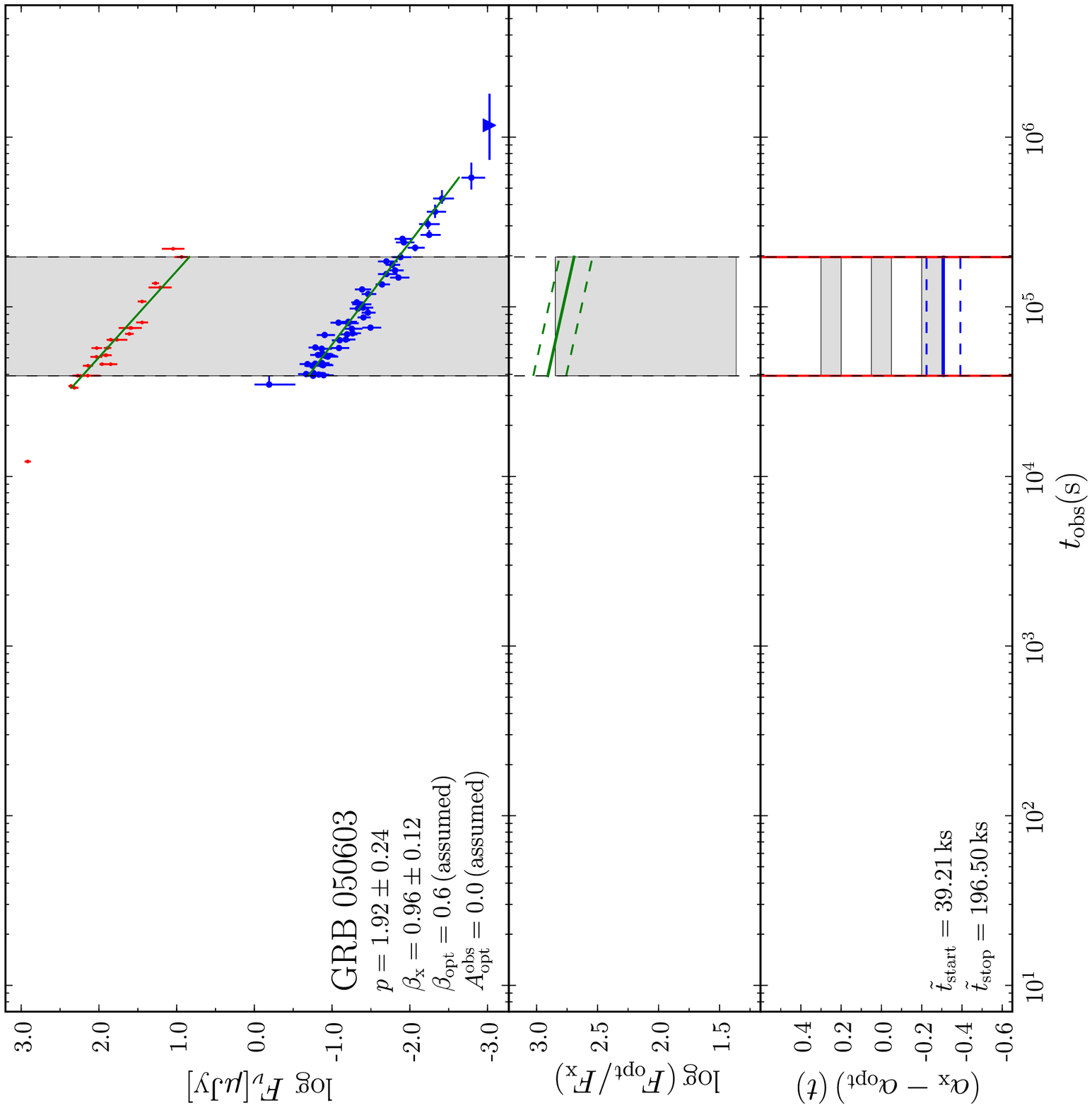}
\includegraphics[bb=44 217 654 835, clip, width=0.5\textwidth, angle=-90]{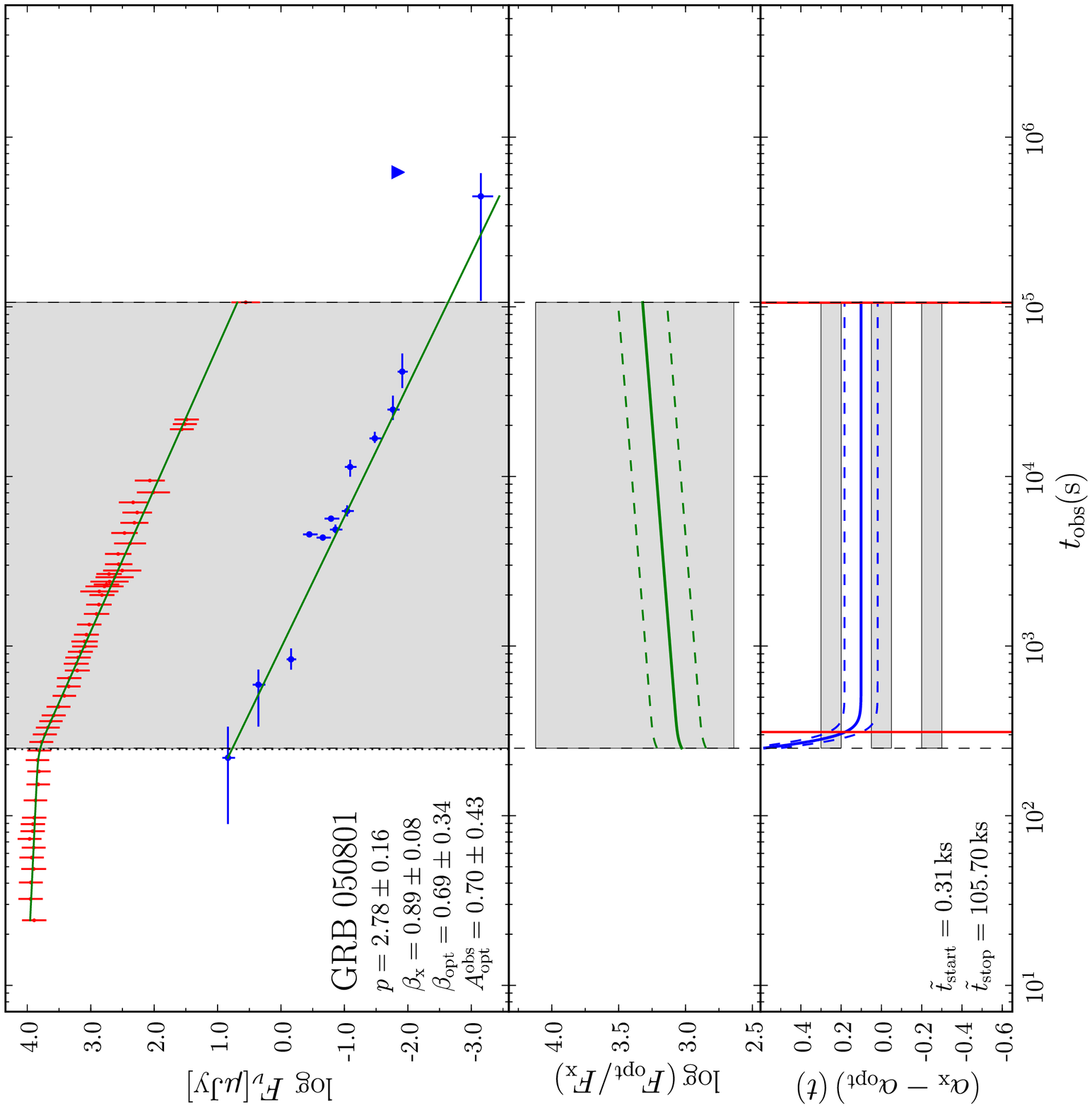}
\includegraphics[bb=44 217 654 835, clip, width=0.5\textwidth, angle=-90]{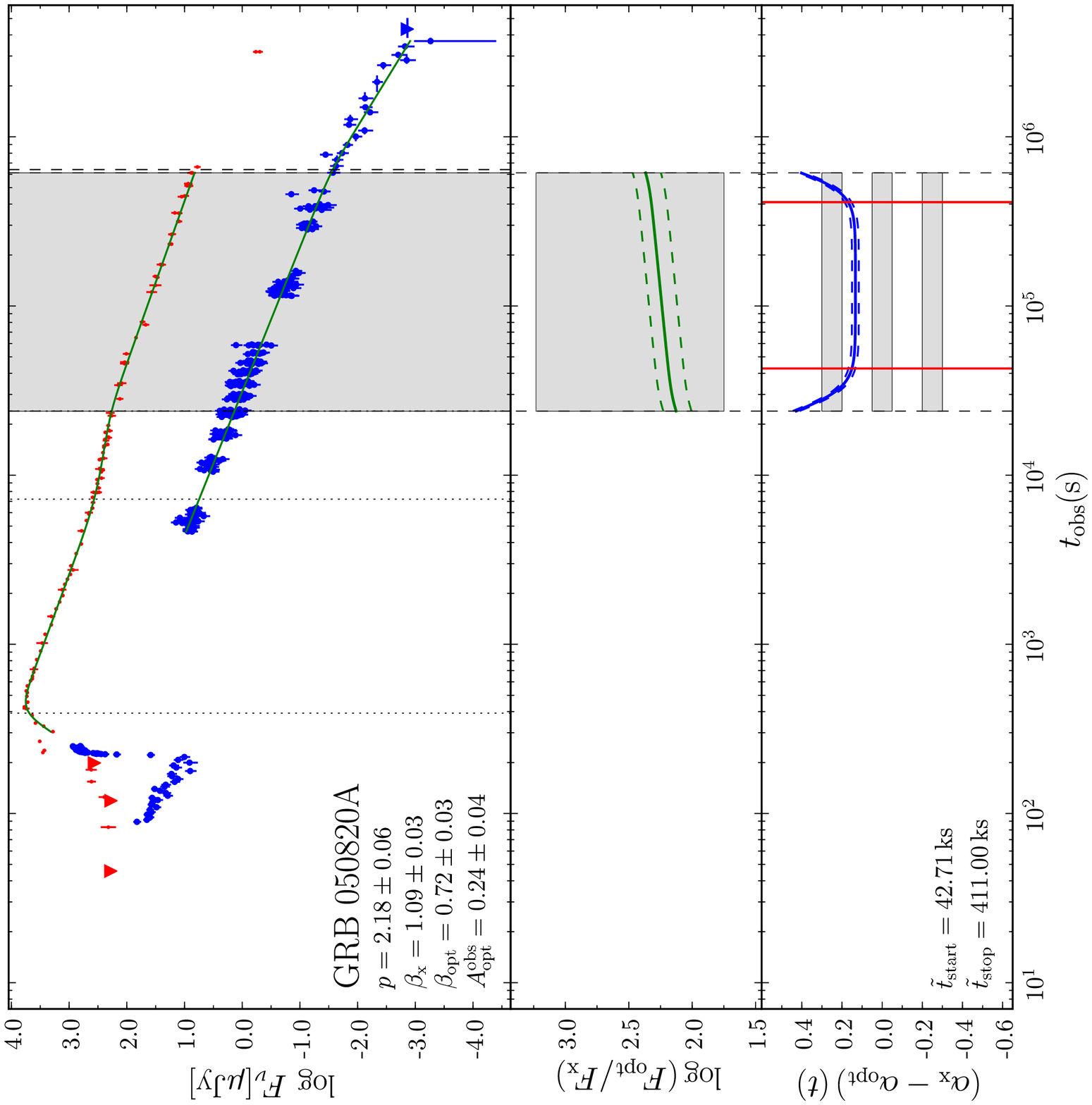}
\includegraphics[bb=44 217 654 835, clip, width=0.5\textwidth, angle=-90]{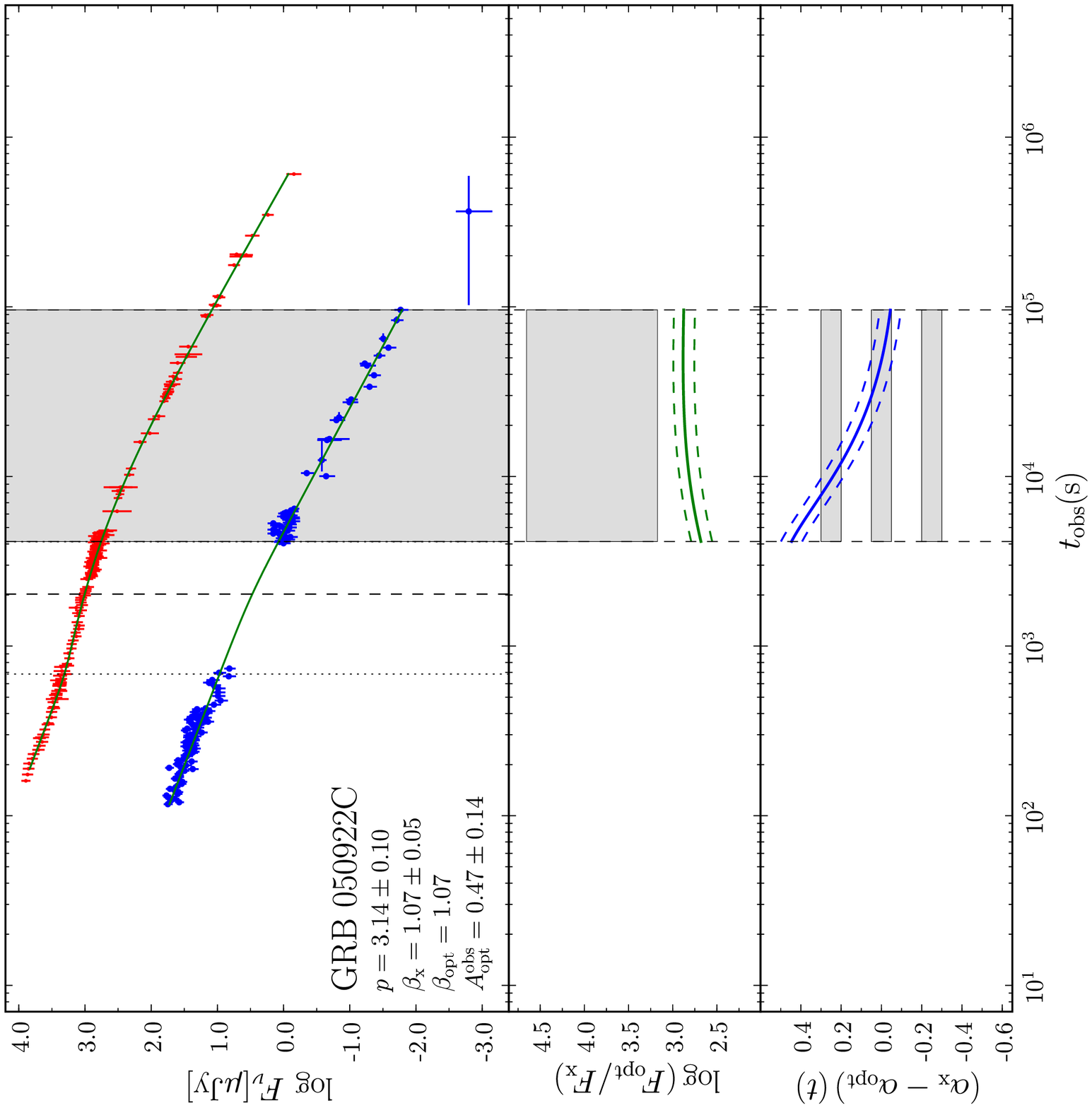}

\caption{Optical and X-ray afterglow light curves of the 27 bursts that entered our sample.
	\textit{Upper panel}:
	The optical data in the $R_c$ band are shown as dots and the X-ray data at 1.73 keV as bigger dots
	with an error bar in time. The light-curve fits are over-plotted. Upper limits are shown as
	downwards-pointing triangles. The grey box is the overlapping time interval of the late-time evolution.
	Vertical dotted and dashed lines indicate breaks in the optical and X-ray band. Information on the
	SEDs are shown in the bottom left (see also Table \ref{tab:sed}). The given extinction, $A^{\rm host} _{\rm opt}$,
	is the observed host-extinction in the $R_c$ band based on the deduced host extinction in the $V$-band, $A^{\rm host} _{V}$.
	Additionally, we deduced the electron index, $p$, from $\beta_{\rm x}$. The electron index is
	either $p=2\beta$ if $\nu_c<\nu_{\rm x}$ or $p=2\beta+1$ if $\nu_c>\nu_{\rm x}$ \citep[e.g.,][]{Zhang2004a}.
	Its error was computed by propagating the uncertainty in $\beta_{\rm x}$.
	\textit{Middle panel}:
	The flux density ratio between the optical and X-ray afterglow is shown as a solid line and its error as a
	dashed line for the shared time interval of the late-time evolution. The grey box represents the allowed
	parameter space of the flux density ratio (Table \ref{tab:predictions}). The upper boundary is the expected
	flux density ratio for $\nu_c\leq \nu_{\rm opt}$, while the lower one shows the expected ratio for
	$\nu_c \geq \nu_{\rm x}$. If the cooling break is in between the optical and the X-ray bands, the expected
	flux-density ratio lies be in between these boundaries. The expected flux density ratio depends on the electron
	index. Not all bursts could be corrected for host extinction. The error on the electron index was neither
	propagated into the error of the expected nor of the observed flux-density ratio.
	\textit{Lower panel}: 
	The first logarithmic derivative of the flux-density ratio, $\left(\alpha_{\rm x}-\alpha_{\rm opt}\right)(t)$,
	is shown as a solid curve and its error is plotted as a dashed line. For $t/t_{\rm break} \ncong1$, the  
	first logarithmic derivative is identical to the difference in the decay slopes obtained from the light-curve fit
	(asymptotic values). Usually breaks in the light curves tend to be smooth instead of sharp. Because of this, the first
	logarithmic derivative deviates from the asymptotic value close to a break depending on the smoothness of the break.
	Two solid lines are plotted to highlight the time interval when the asymptotic decay slopes were reached within $1\sigma$.
	The precise values are shown on the left and in Table \ref{tab:radii}. 	Within $3\sigma$, the asymptotic difference
	in the decay slopes agrees either with $+1/4$, 0, $-1/4$ depending on the spectral and dynamical regime
	and the circumburst density profile. Furthermore, an envelope is drawn around expected values, $+1/4$, $0$, $-1/4$,
	with a width of 0.1 to guide the eye.
	}
\label{fig:sample}
\end{figure*}
\begin{figure*}
\includegraphics[bb=44 217 654 835, clip, width=0.5\textwidth, angle=-90]{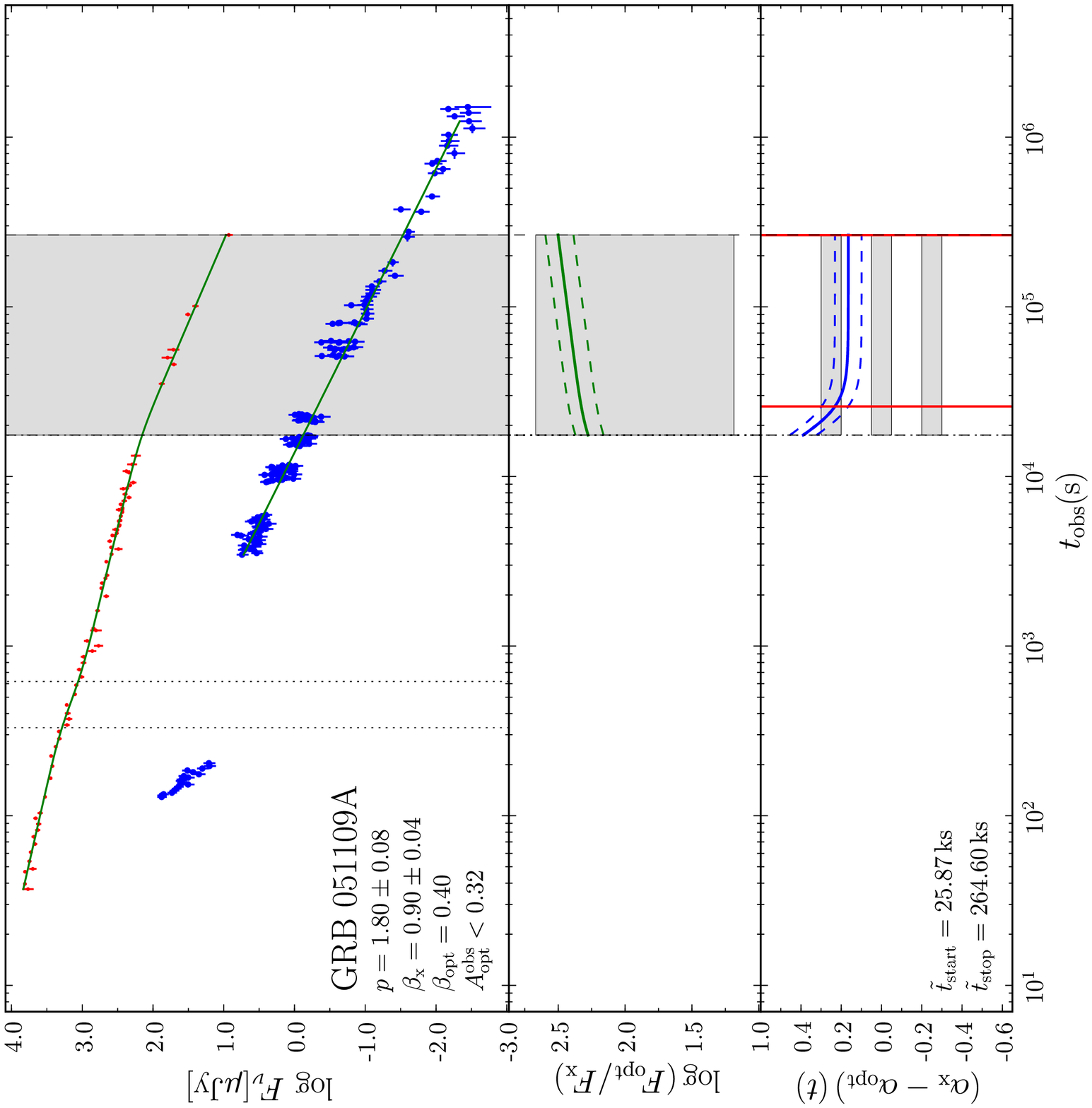}
\includegraphics[bb=44 217 654 835, clip, width=0.5\textwidth, angle=-90]{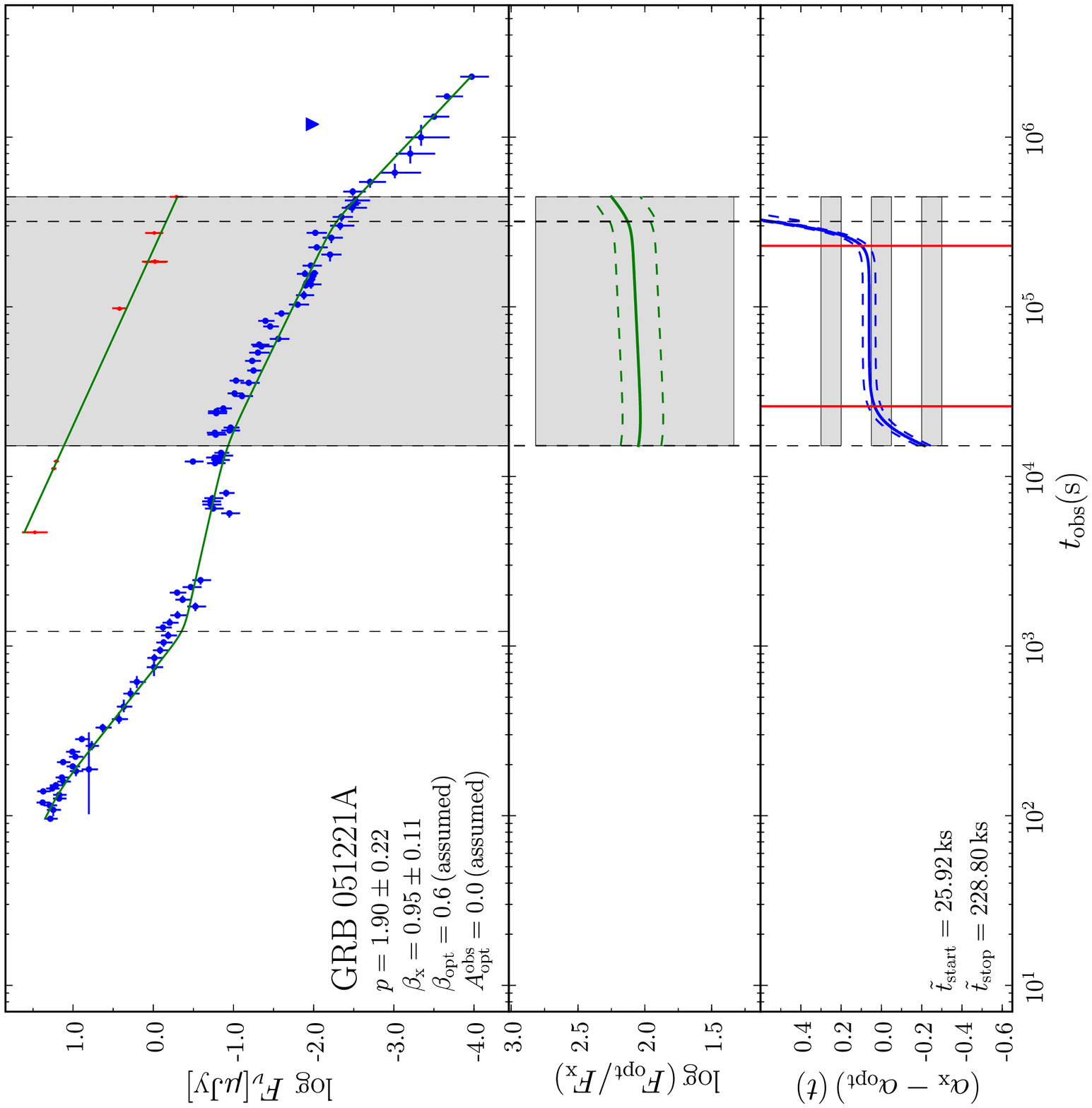}
\includegraphics[bb=44 217 654 835, clip, width=0.5\textwidth, angle=-90]{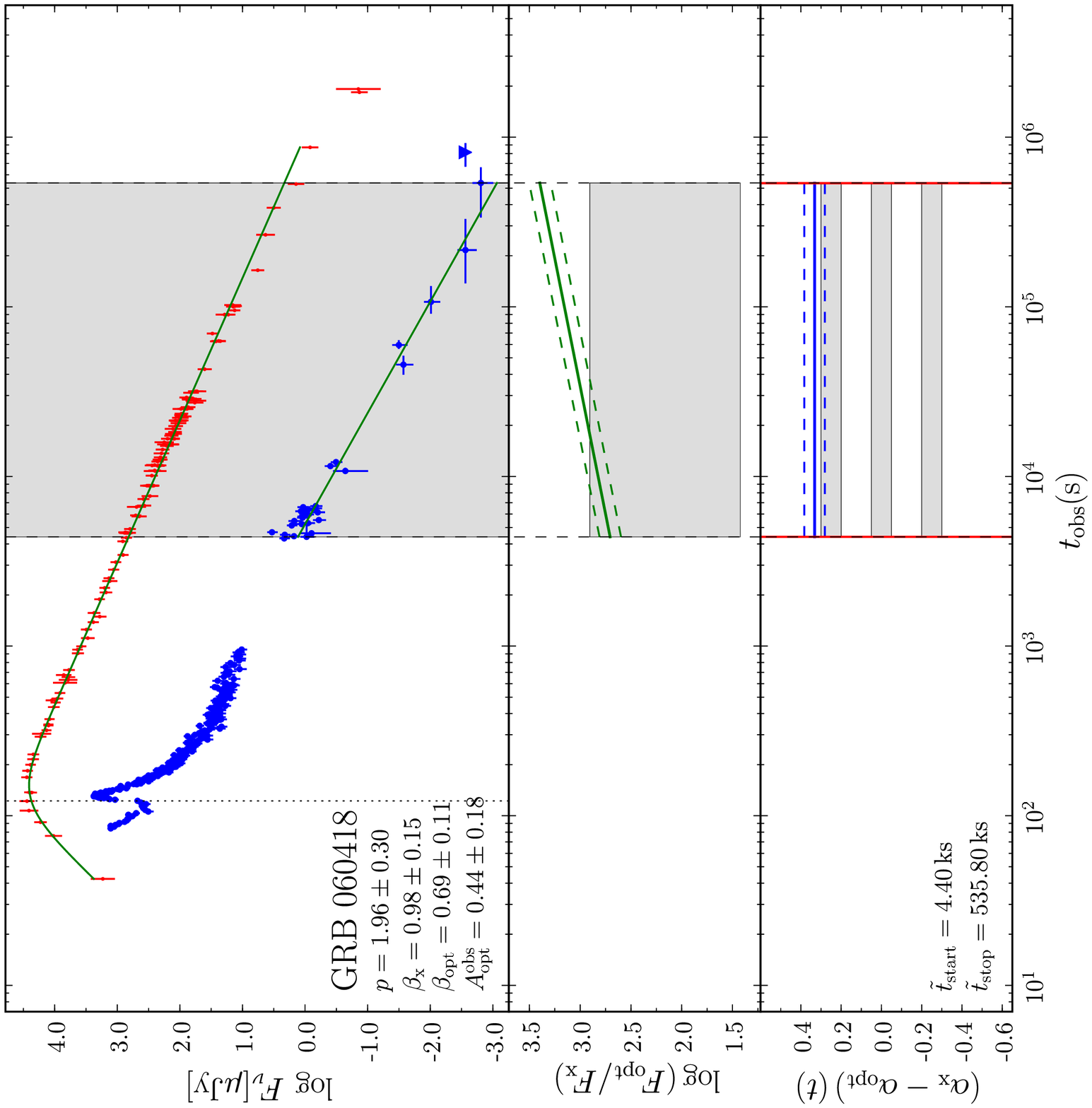}
\includegraphics[bb=44 217 654 835, clip, width=0.5\textwidth, angle=-90]{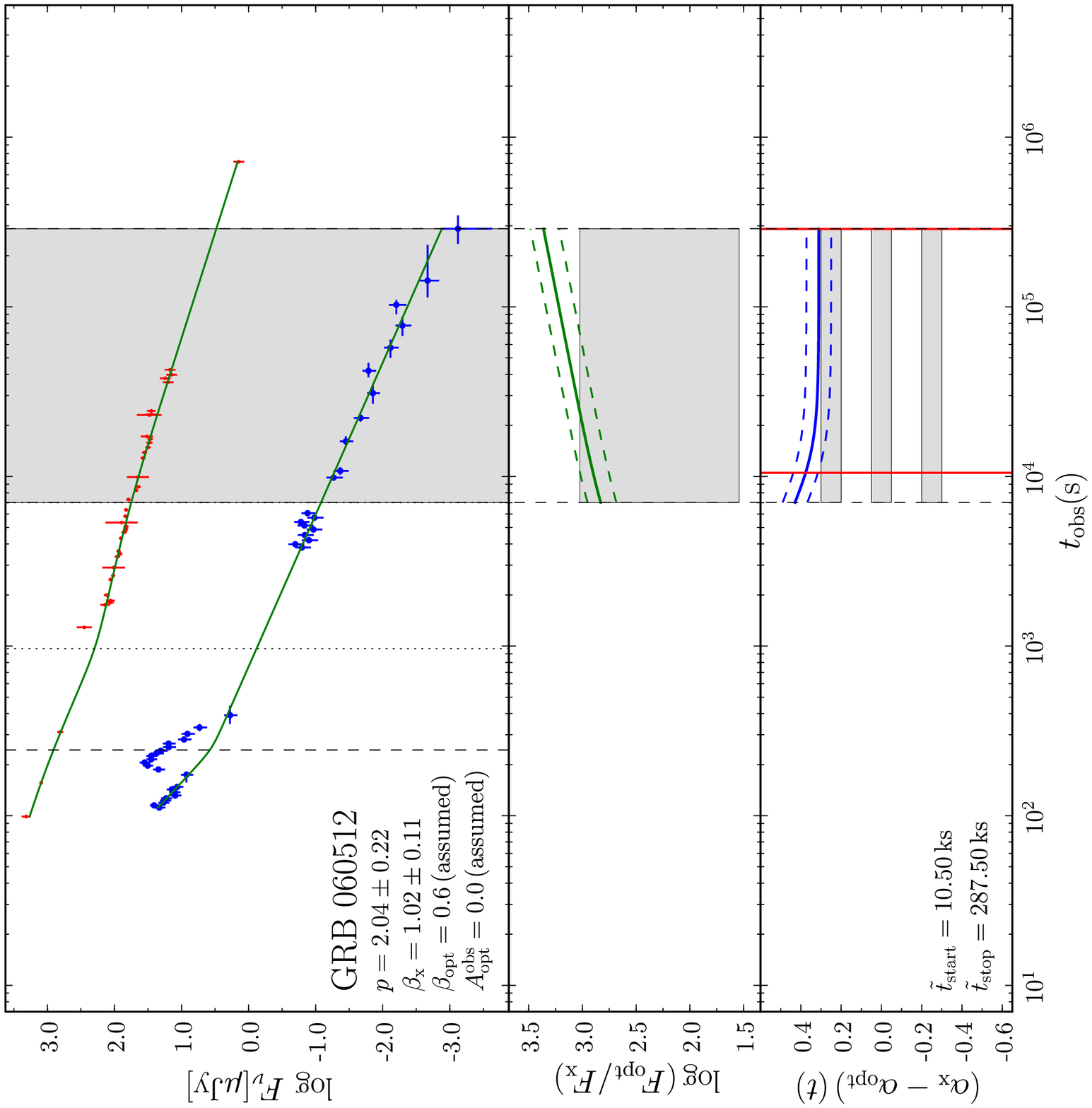}
\hfill\center{Fig. \ref{fig:sample} ---  continued}
\end{figure*}
\begin{figure*}
\includegraphics[bb=44 217 654 835, clip, width=0.5\textwidth, angle=-90]{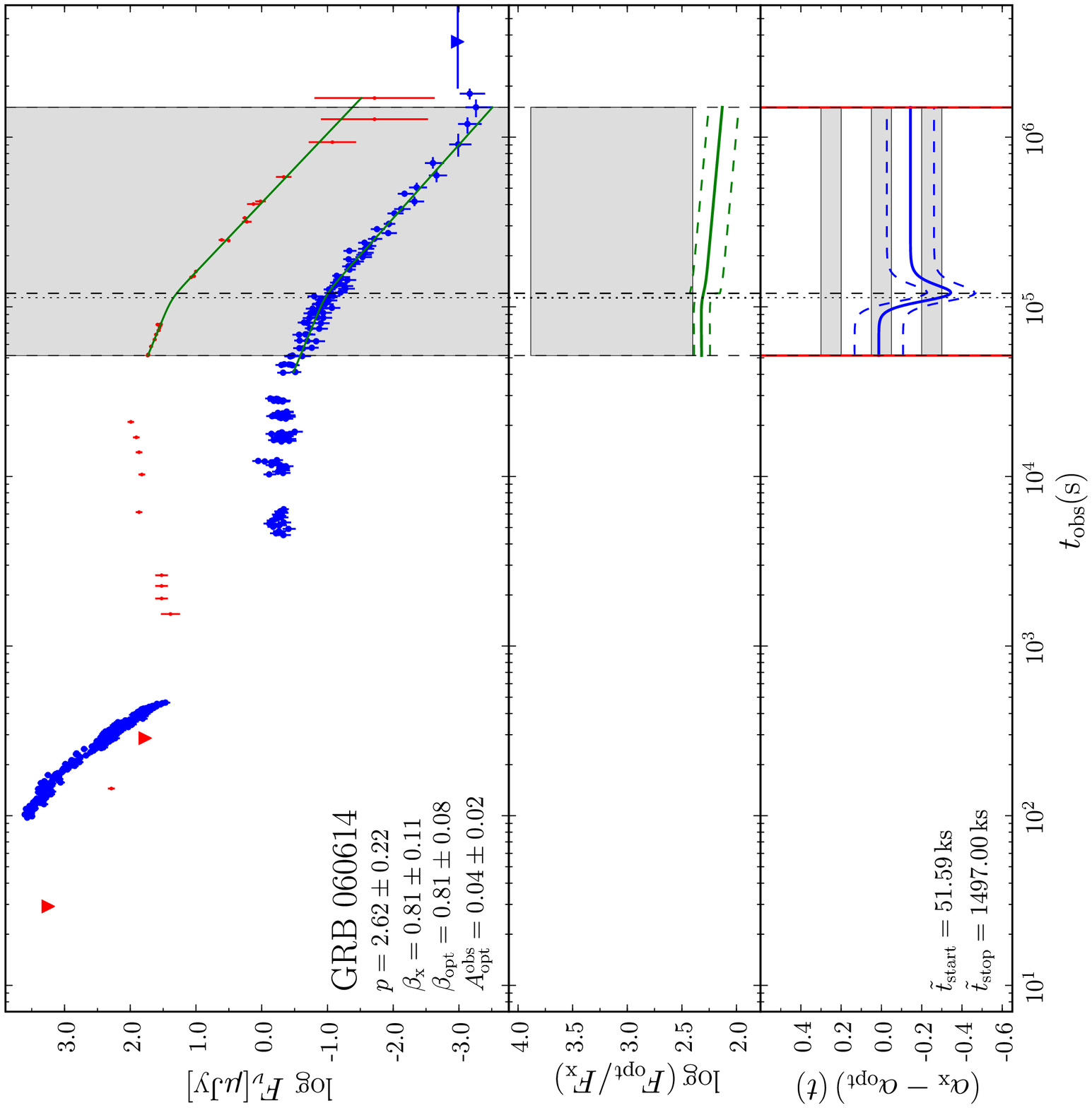}
\includegraphics[bb=44 217 654 835, clip, width=0.5\textwidth, angle=-90]{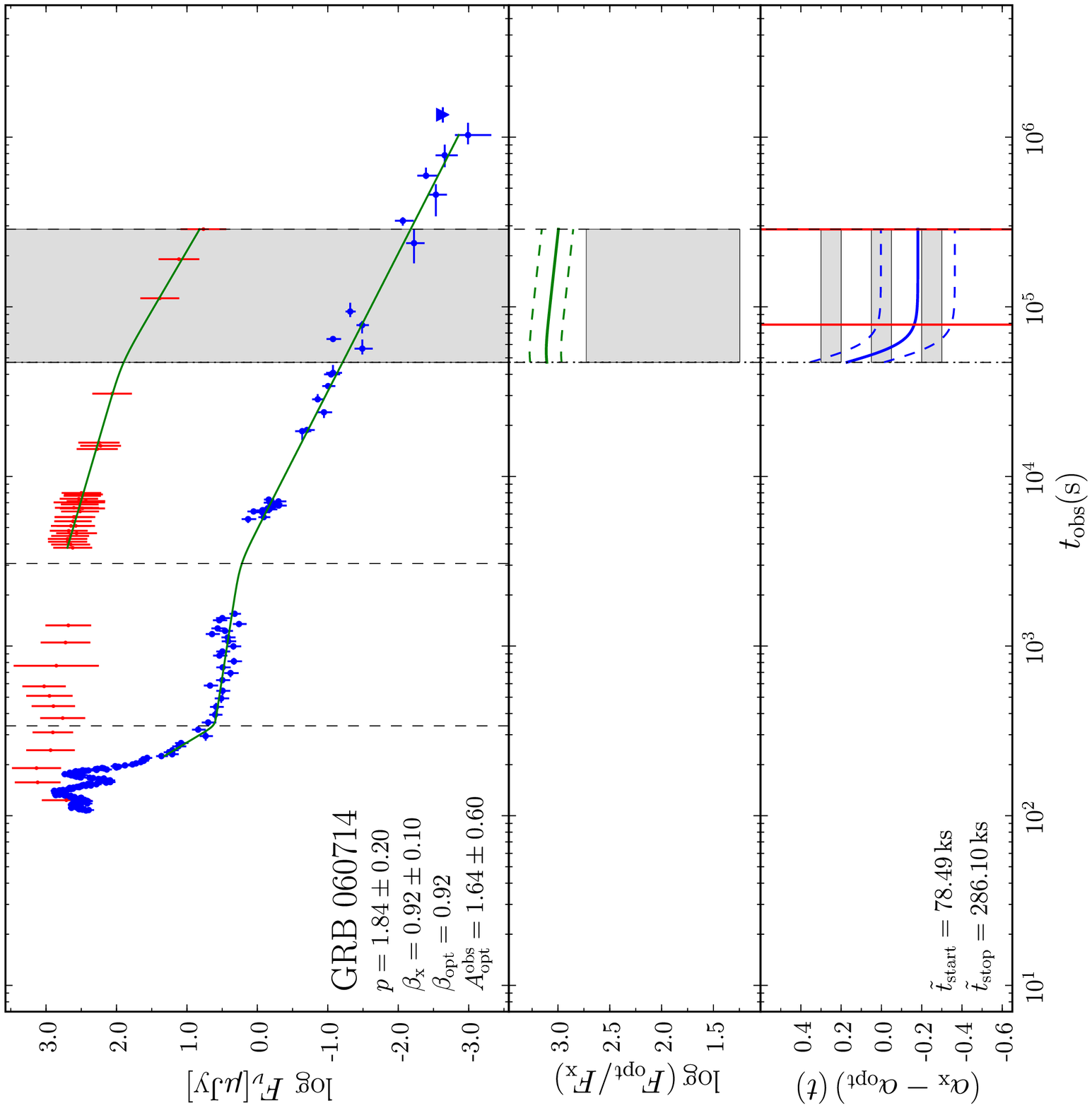}
\includegraphics[bb=44 217 654 835, clip, width=0.5\textwidth, angle=-90]{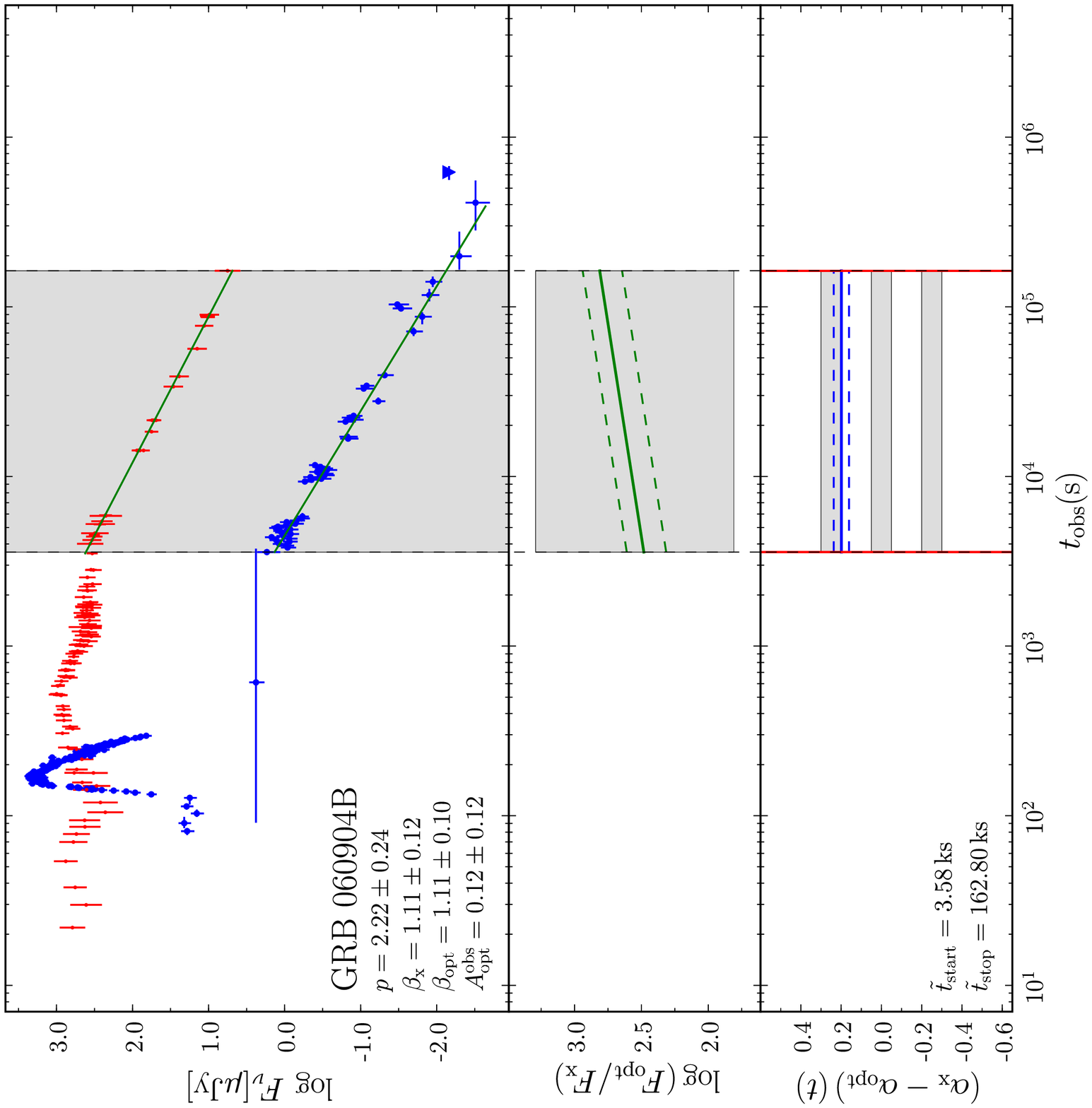}
\includegraphics[bb=44 217 654 835, clip, width=0.5\textwidth, angle=-90]{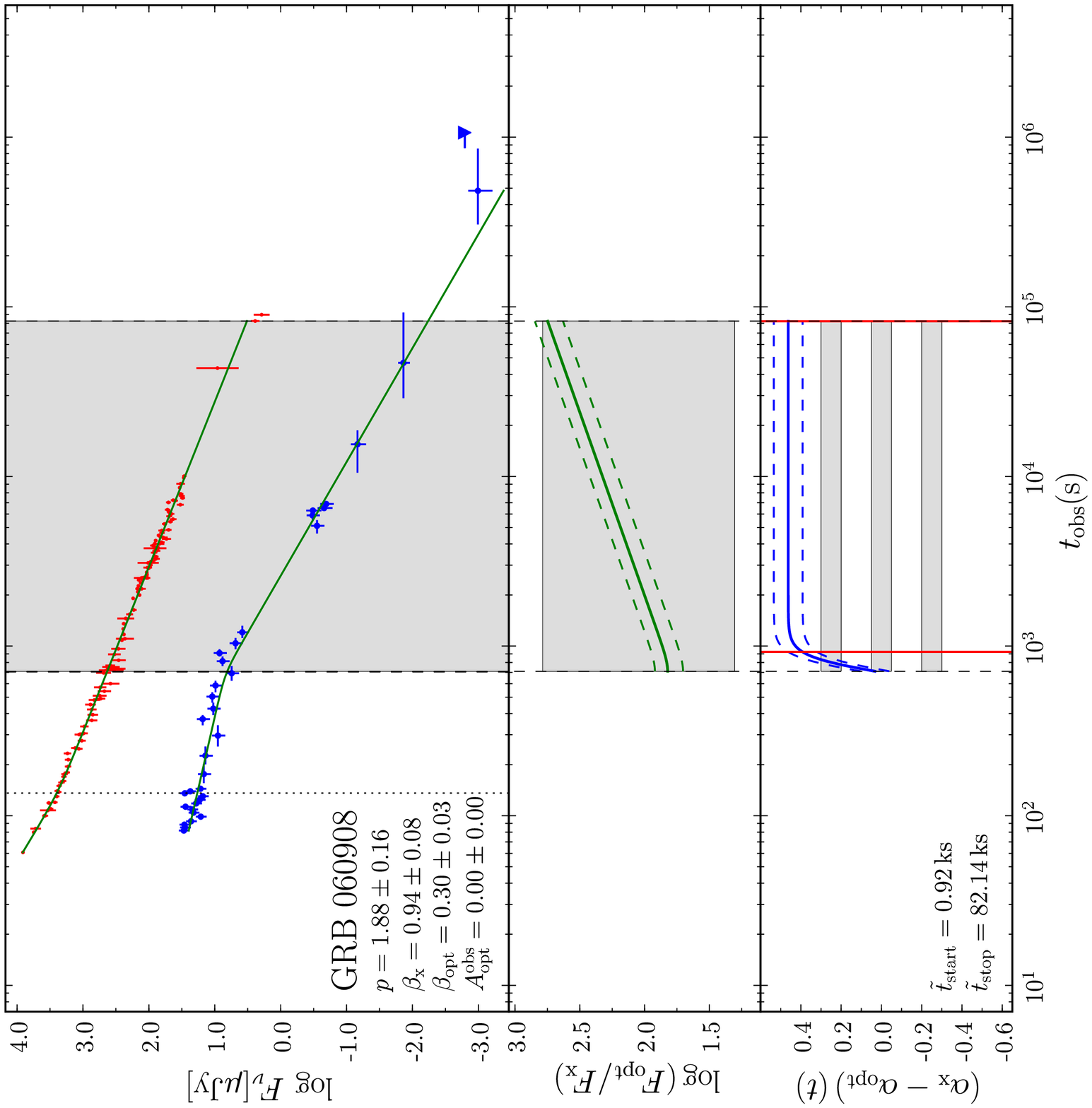}
\hfill\center{Fig. \ref{fig:sample} ---  continued}
\end{figure*}
\begin{figure*}
\includegraphics[bb=44 217 654 835, clip, width=0.50\textwidth, angle=-90]{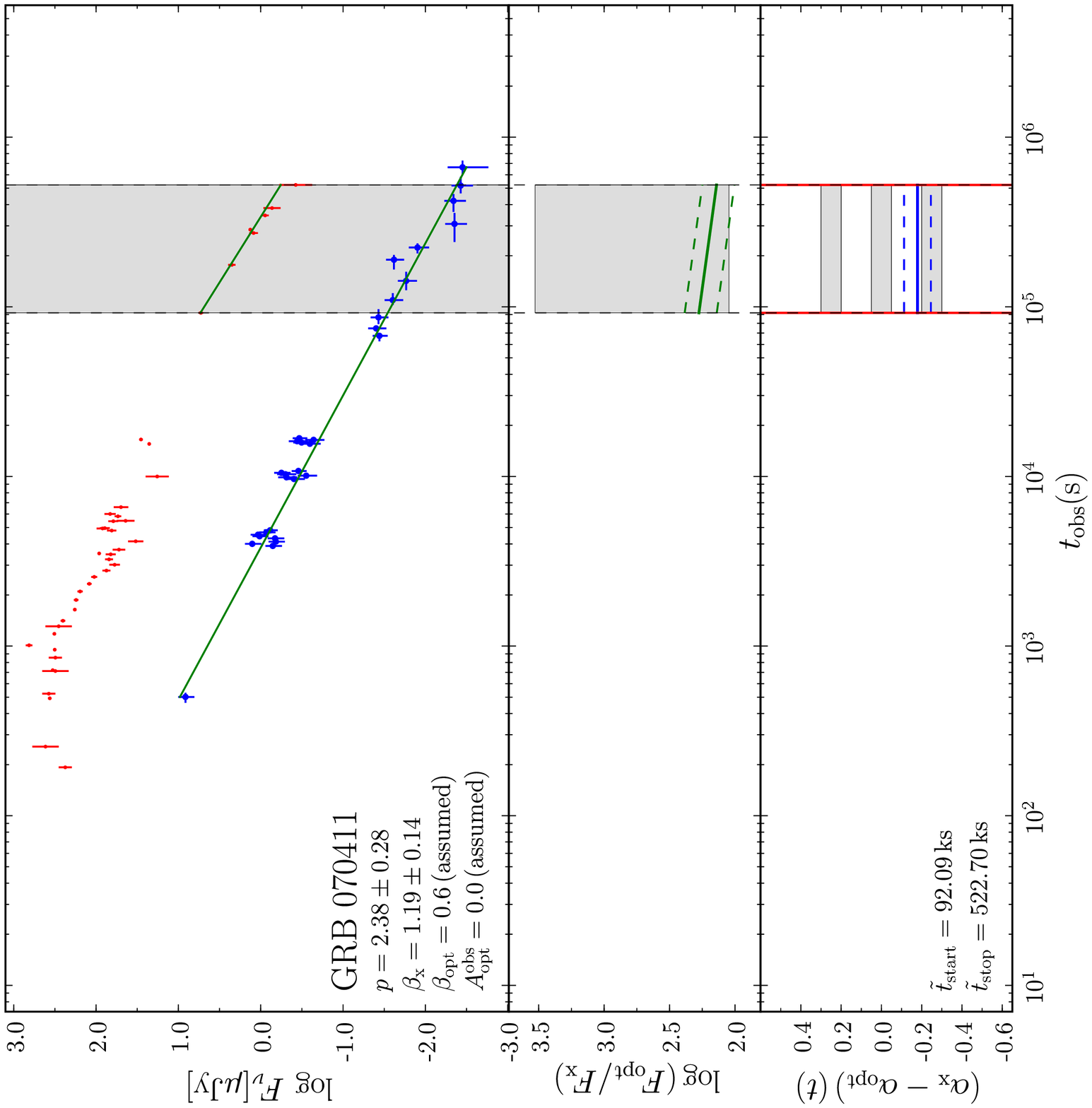}
\includegraphics[bb=44 217 654 835, clip, width=0.50\textwidth, angle=-90]{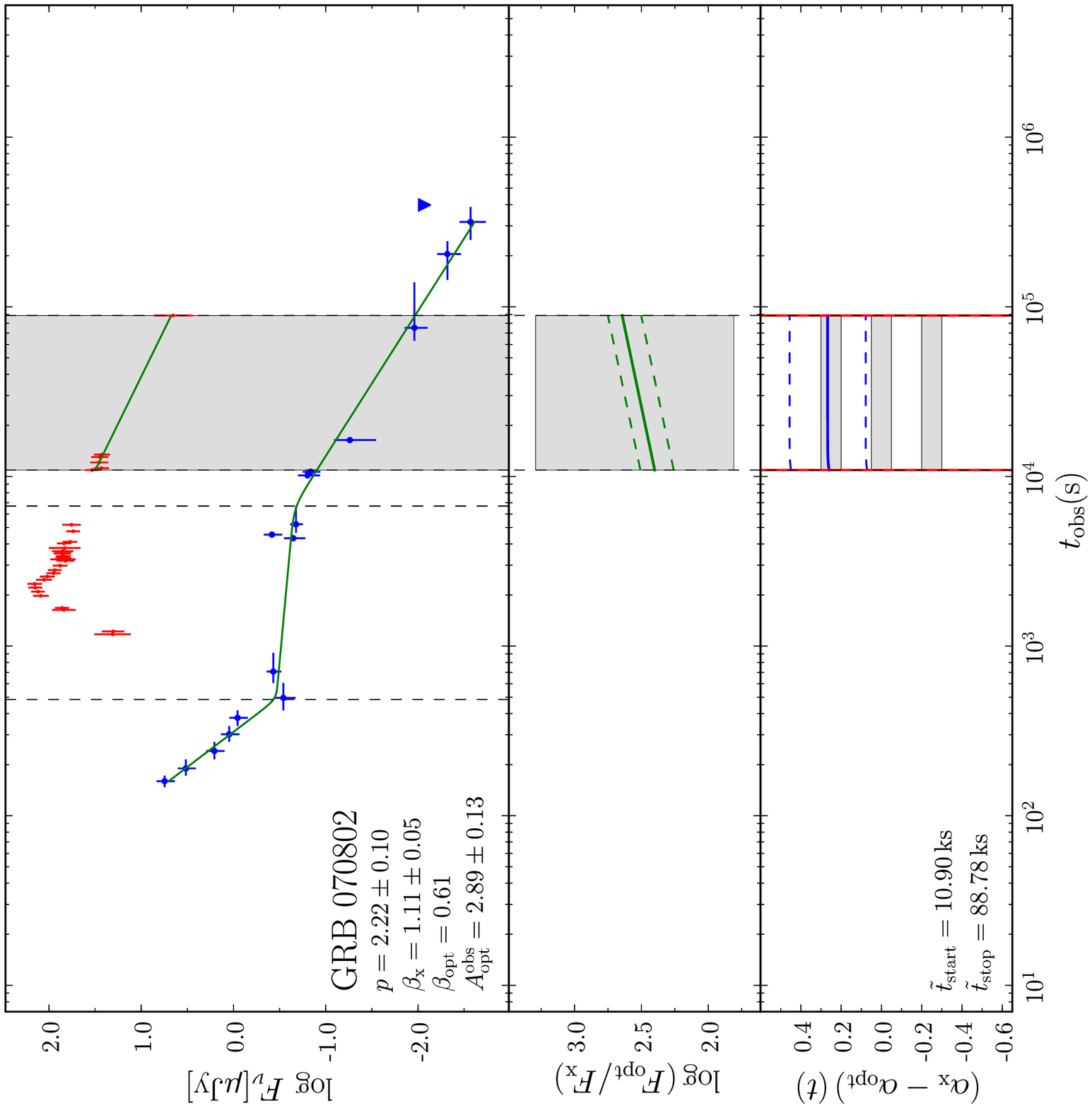}
\includegraphics[bb=44 217 654 835, clip, width=0.50\textwidth, angle=-90]{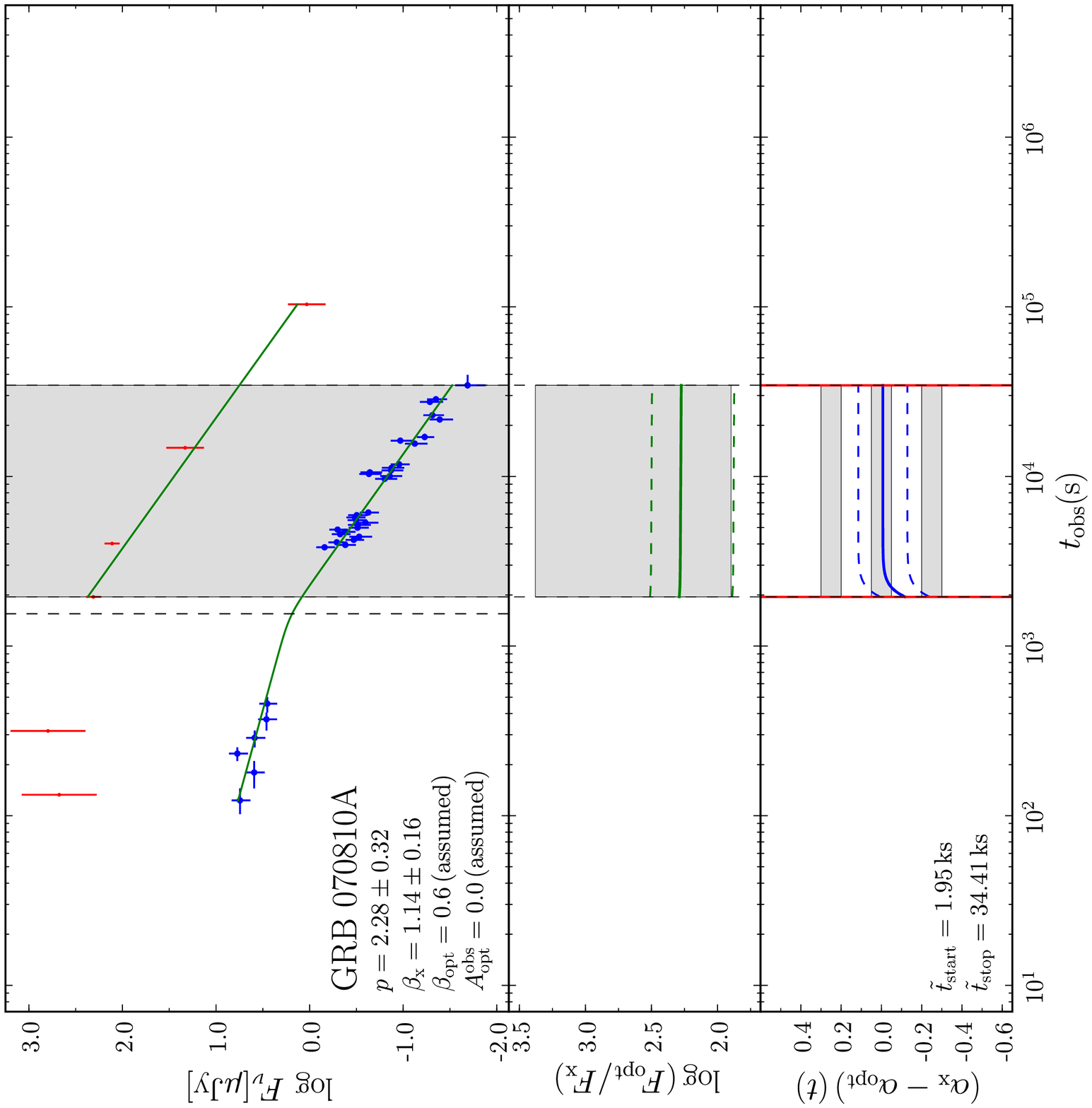}
\includegraphics[bb=44 217 654 835, clip, width=0.50\textwidth, angle=-90]{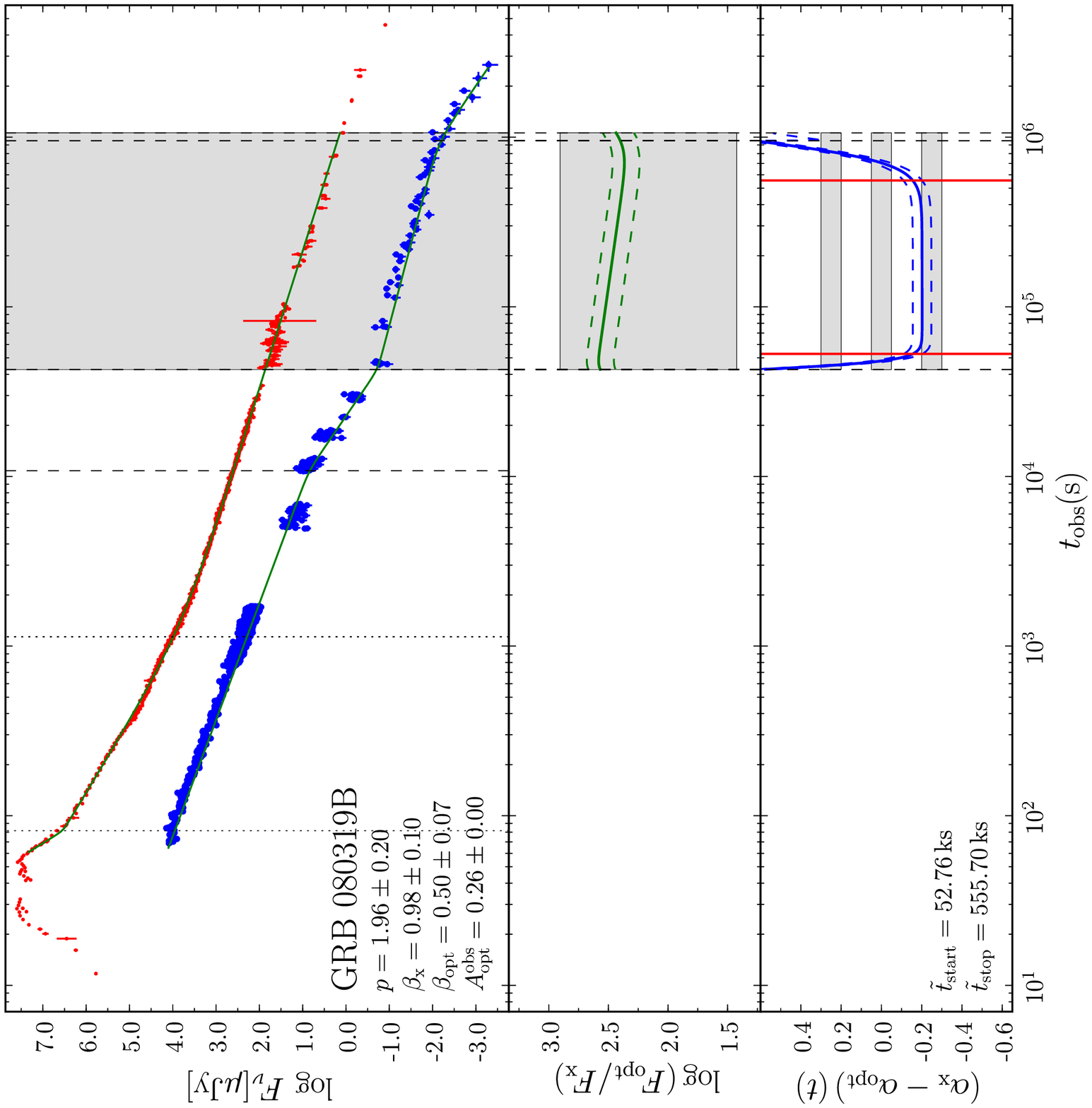}

\hfill\center{Fig. \ref{fig:sample} ---  continued}
\end{figure*}
\begin{figure*}
\includegraphics[bb=44 217 654 835, clip, width=0.50\textwidth, angle=-90]{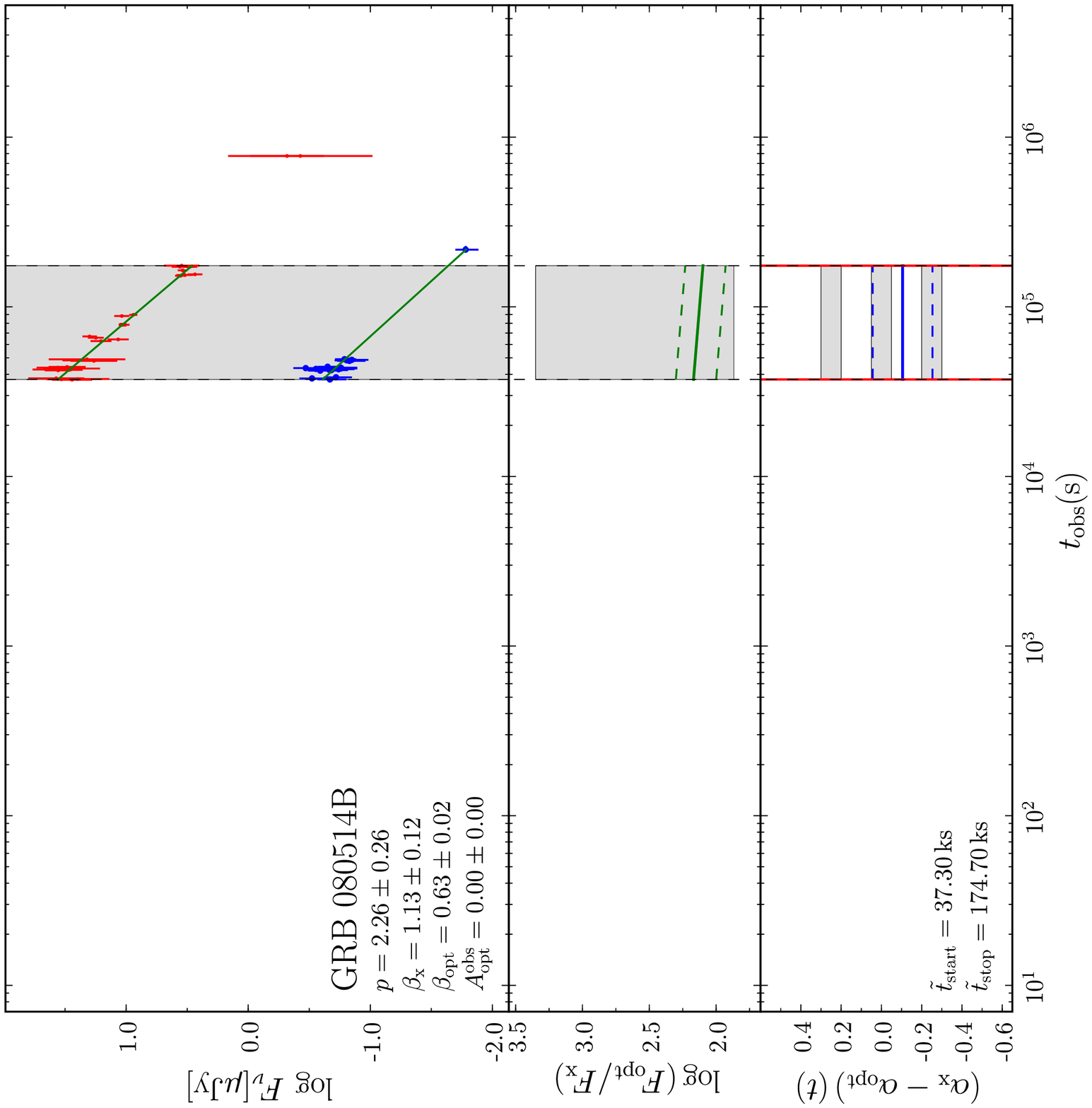}
\includegraphics[bb=44 217 654 835, clip, width=0.50\textwidth, angle=-90]{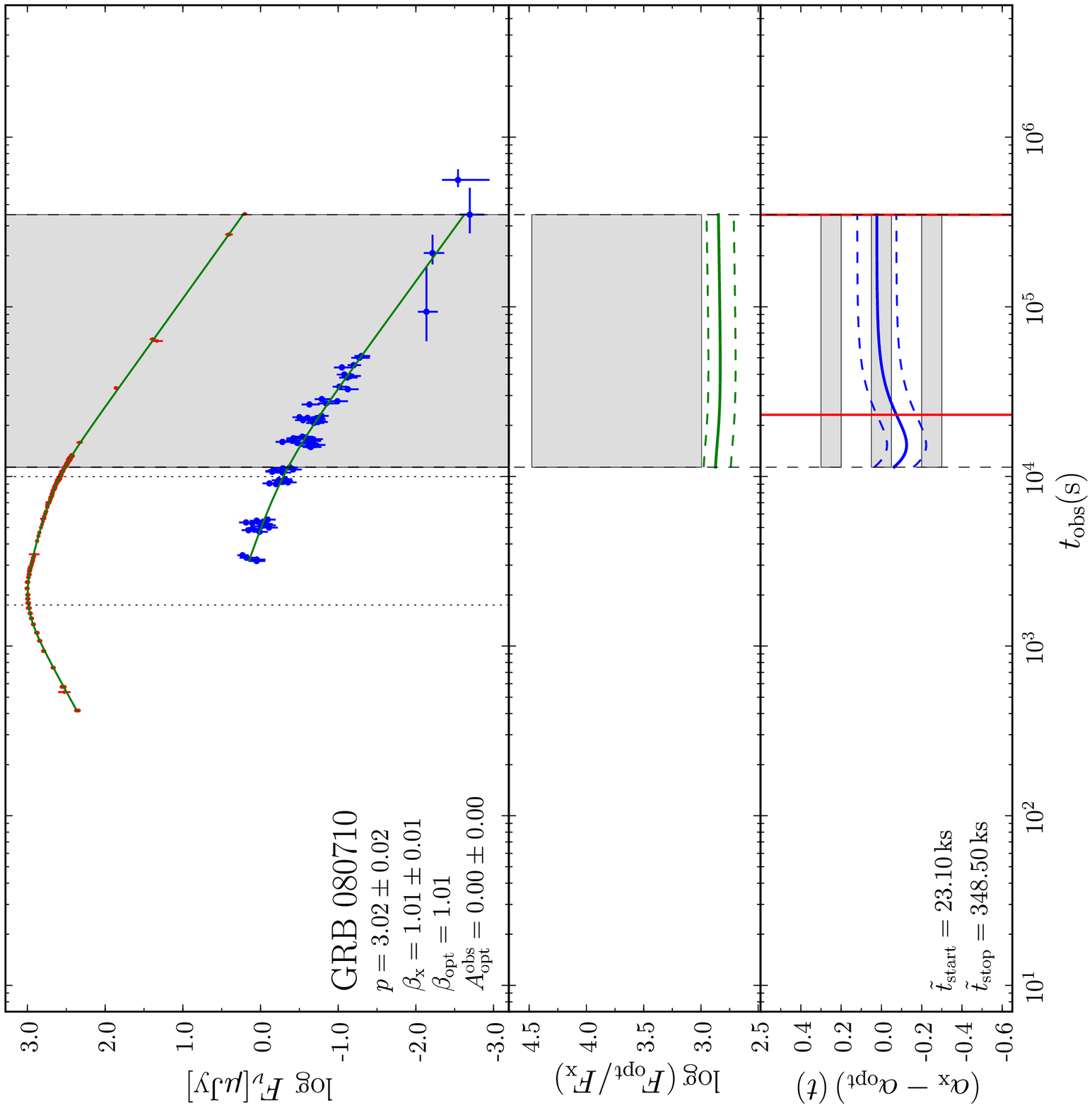}
\includegraphics[bb=44 217 654 835, clip, width=0.50\textwidth, angle=-90]{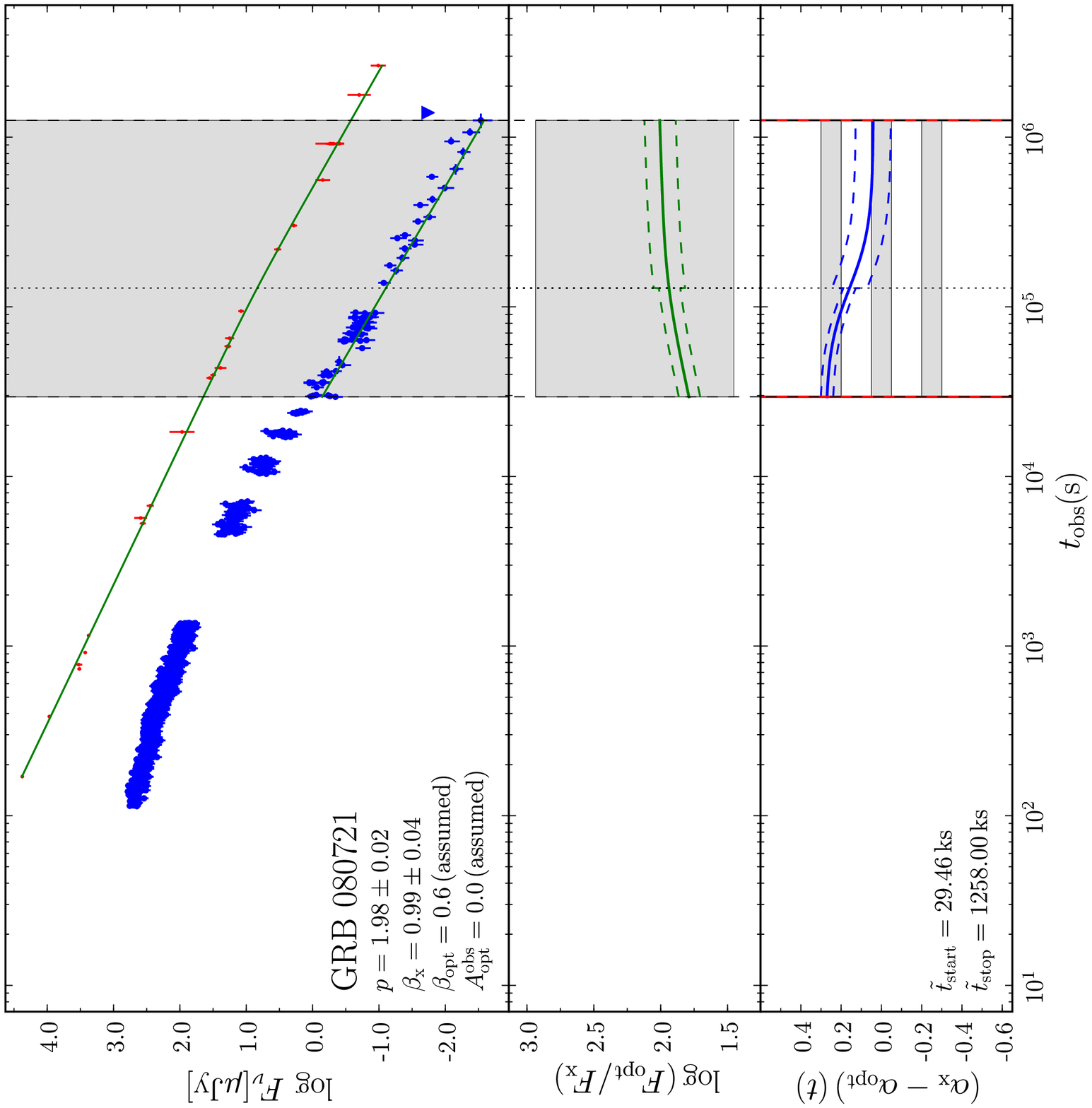}
\includegraphics[bb=44 217 654 835, clip, width=0.50\textwidth, angle=-90]{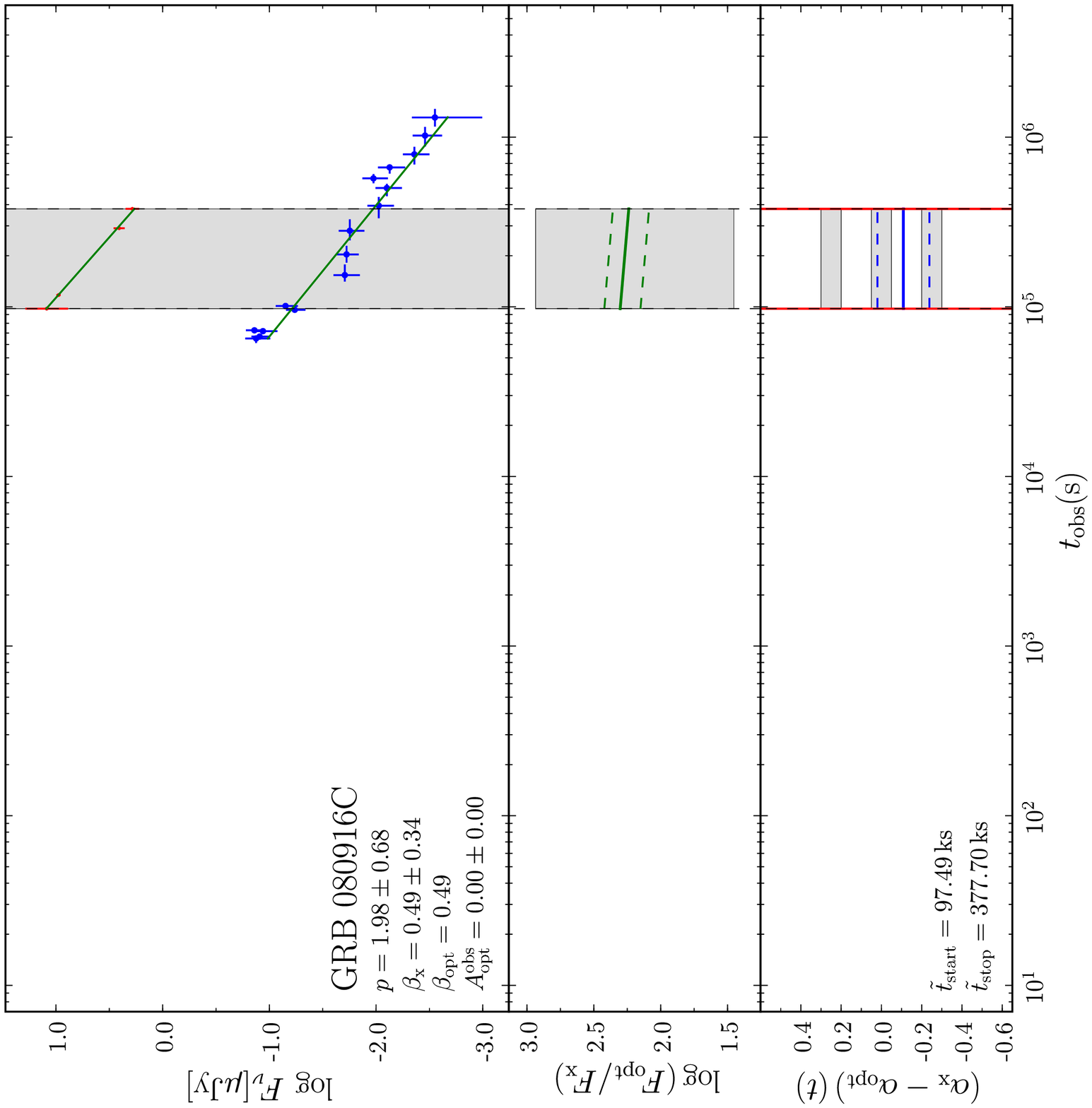}

\hfill\center{Fig. \ref{fig:sample} ---  continued}
\end{figure*}
\begin{figure*}
\includegraphics[bb=44 217 654 835, clip, width=0.50\textwidth, angle=-90]{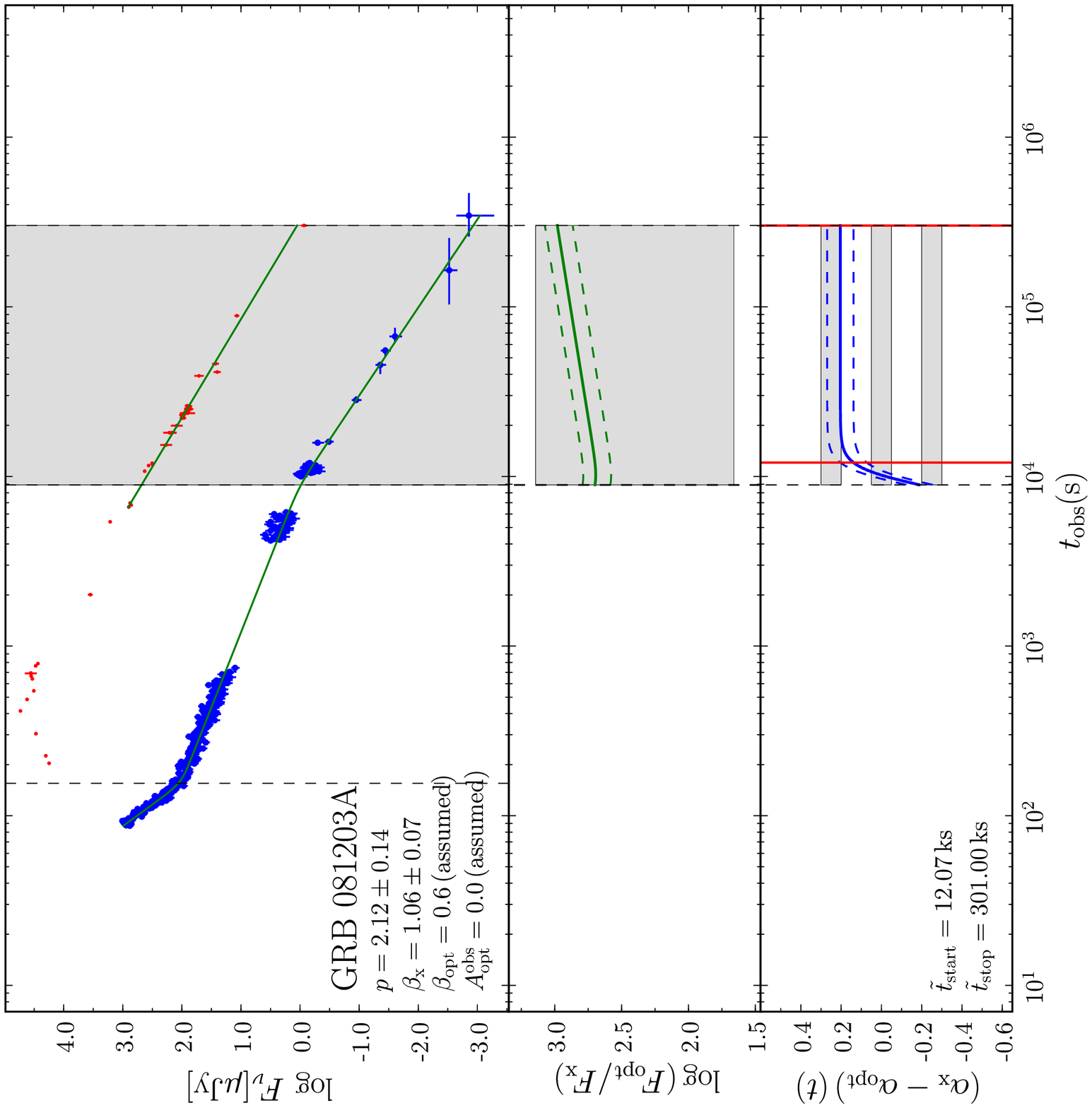}
\includegraphics[bb=44 217 654 835, clip, width=0.50\textwidth, angle=-90]{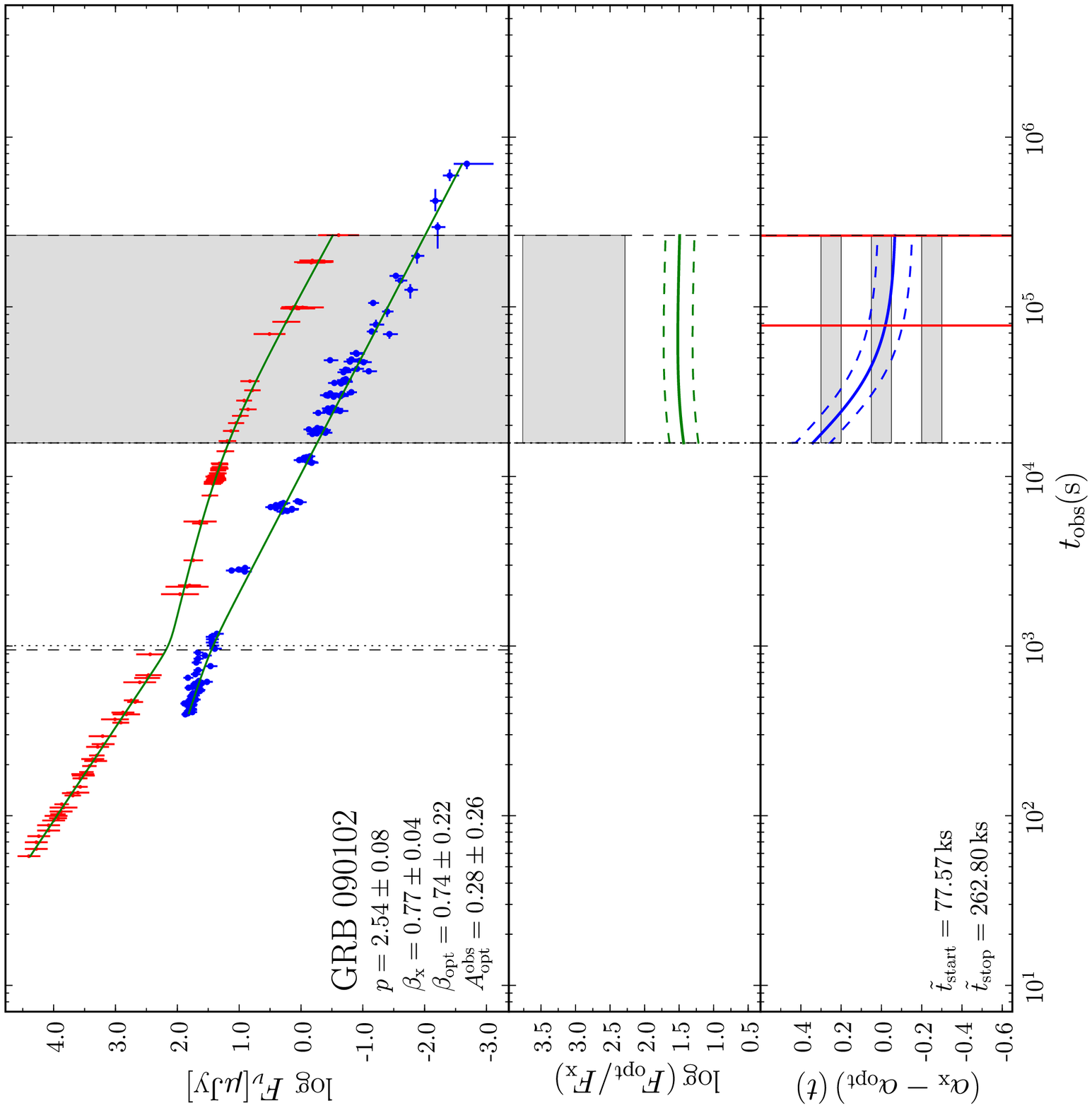}
\includegraphics[bb=44 217 654 835, clip, width=0.50\textwidth, angle=-90]{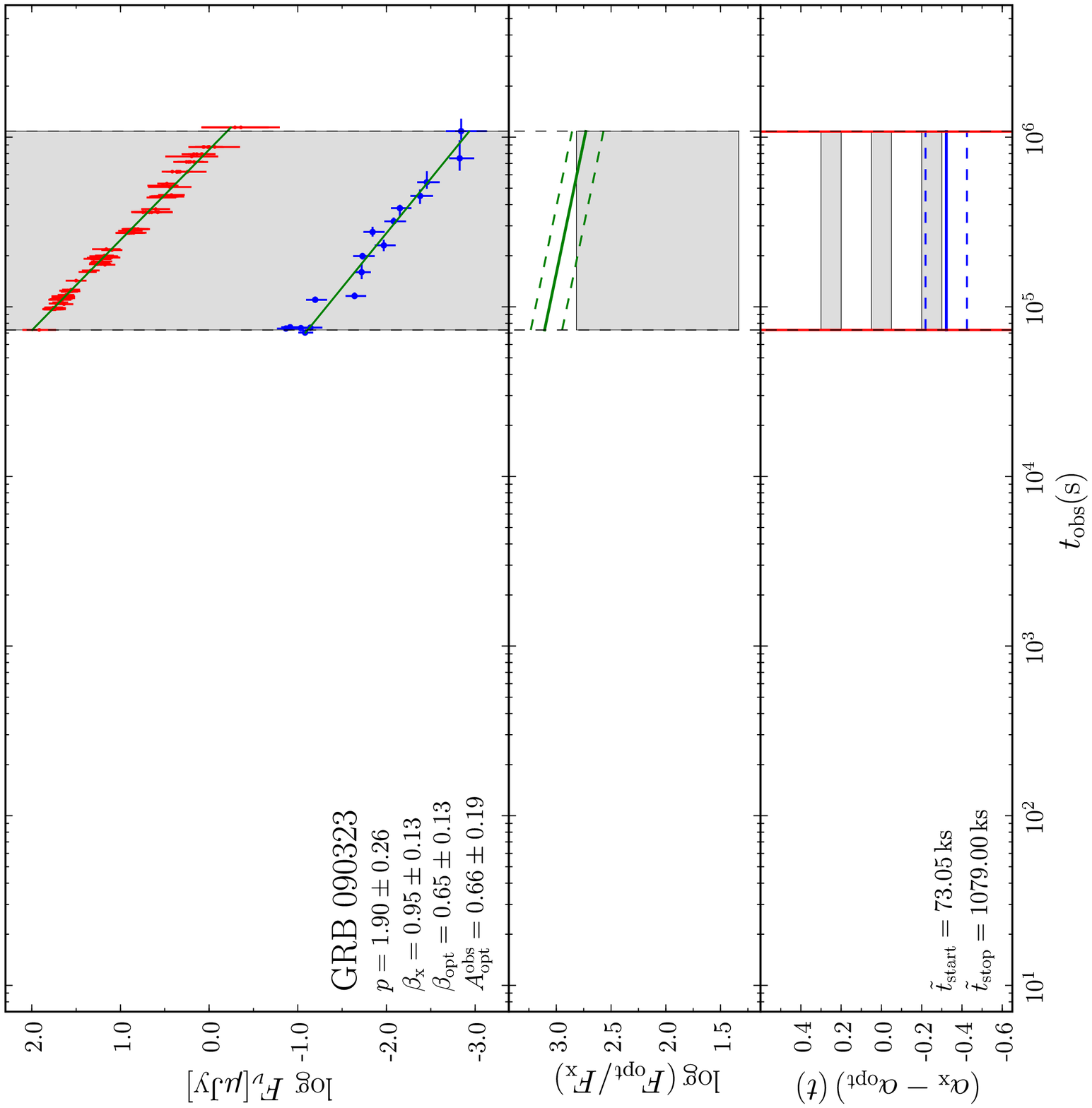}
\includegraphics[bb=44 217 654 835, clip, width=0.50\textwidth, angle=-90]{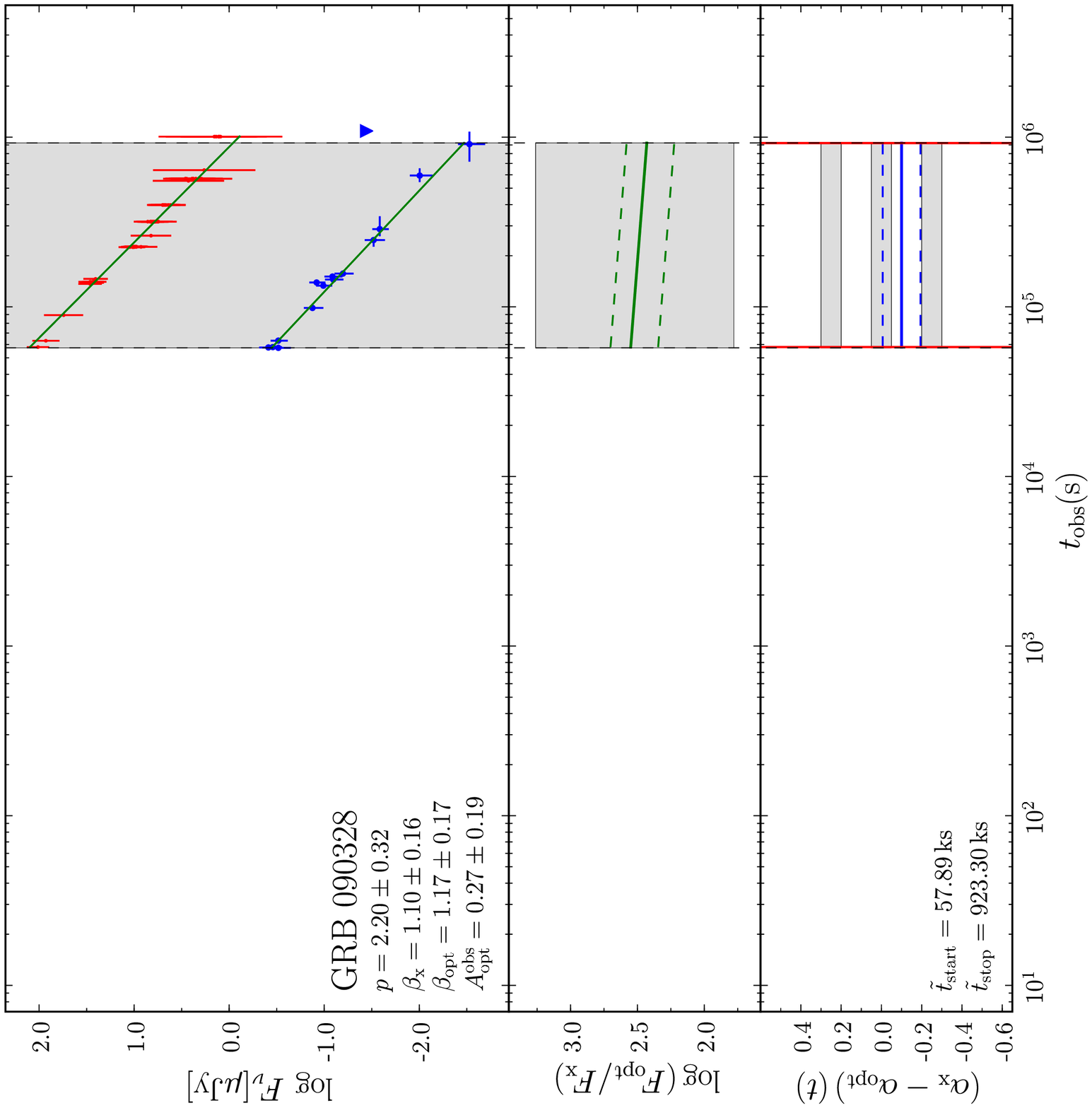}

\hfill\center{Fig. \ref{fig:sample} ---  continued}
\end{figure*}
\begin{figure*}
\includegraphics[bb=44 217 654 835, clip, width=0.50\textwidth, angle=-90]{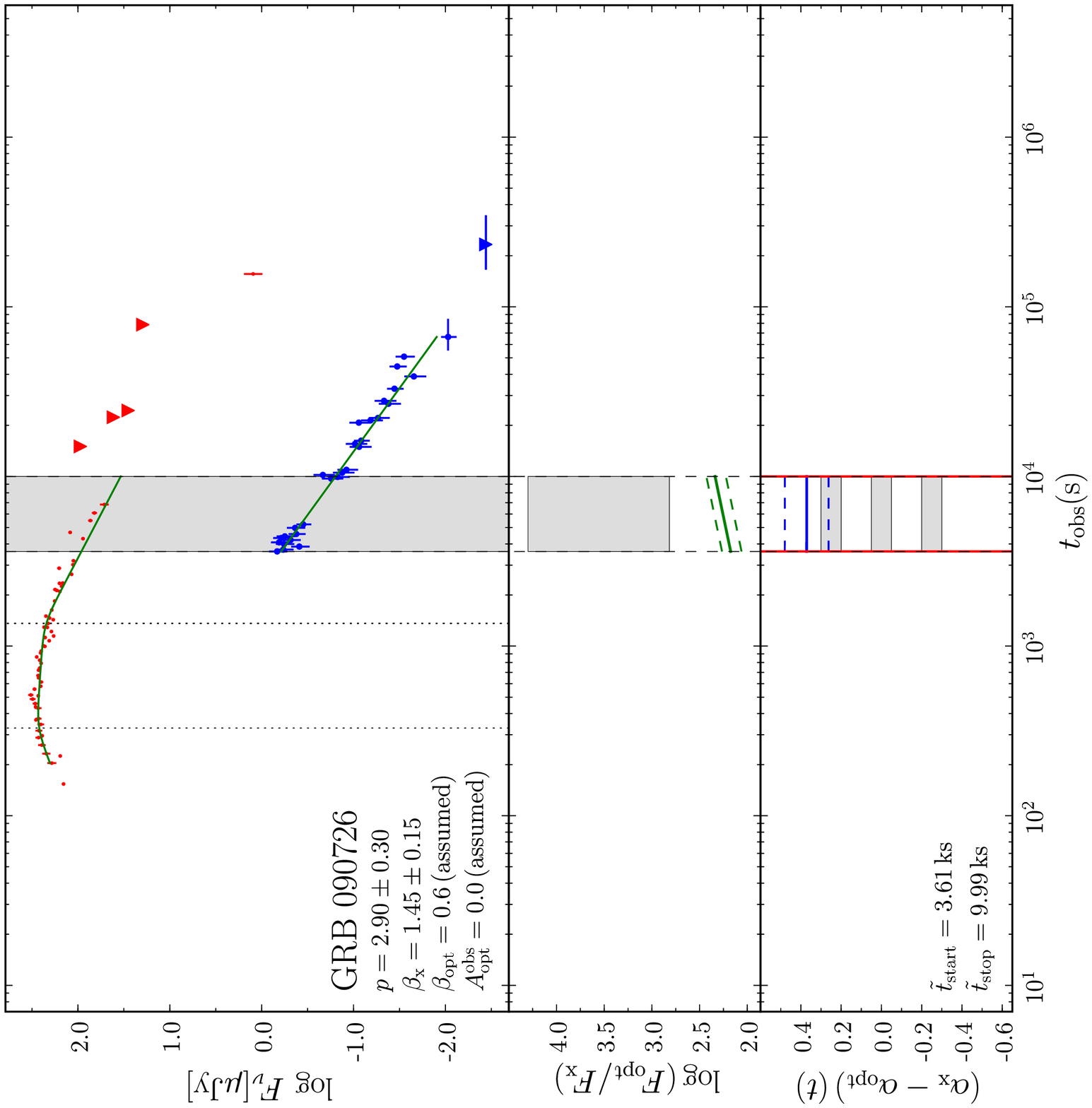}
\includegraphics[bb=44 217 654 835, clip, width=0.50\textwidth, angle=-90]{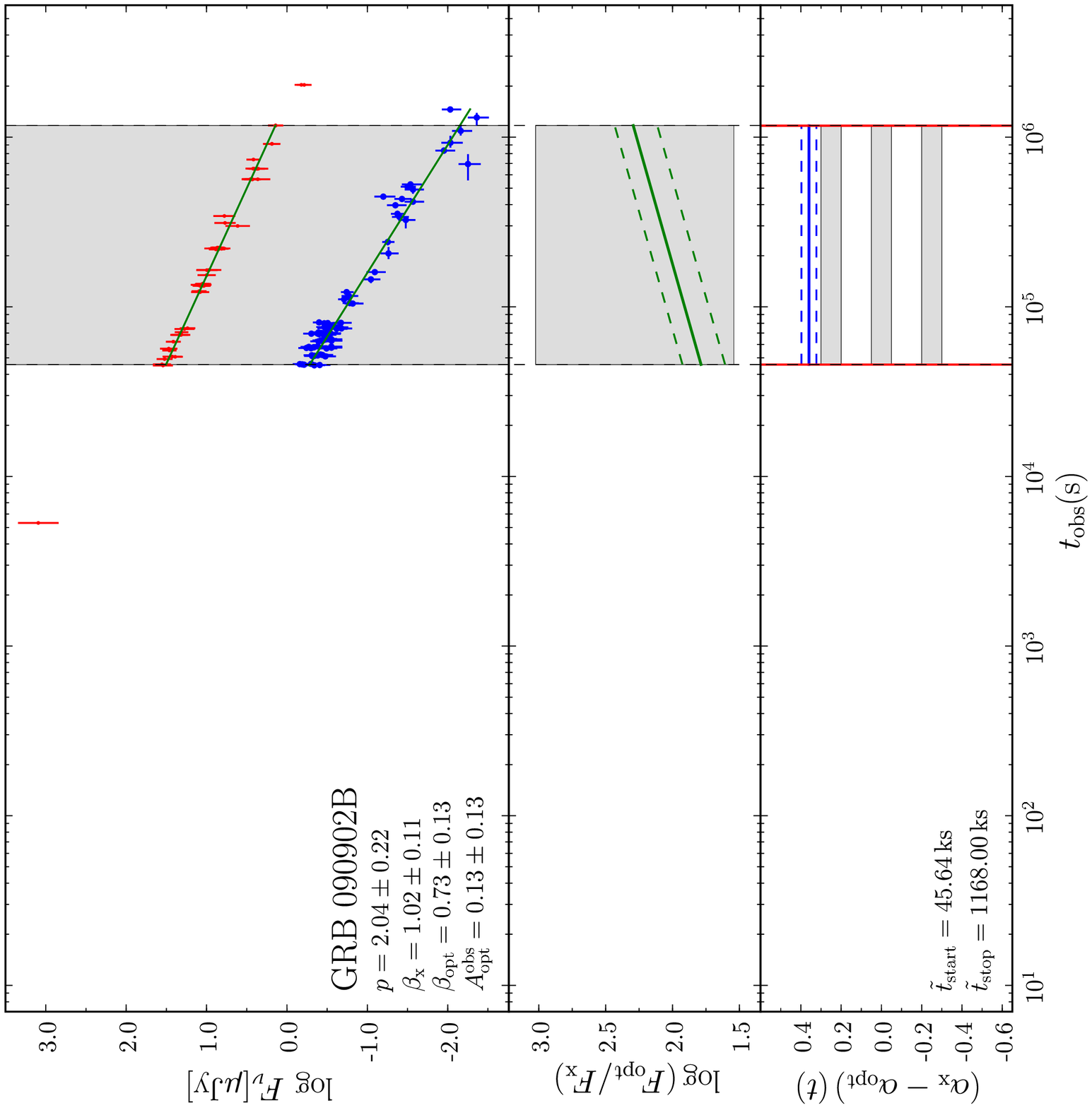}
\includegraphics[bb=44 217 654 835, clip, width=0.50\textwidth, angle=-90]{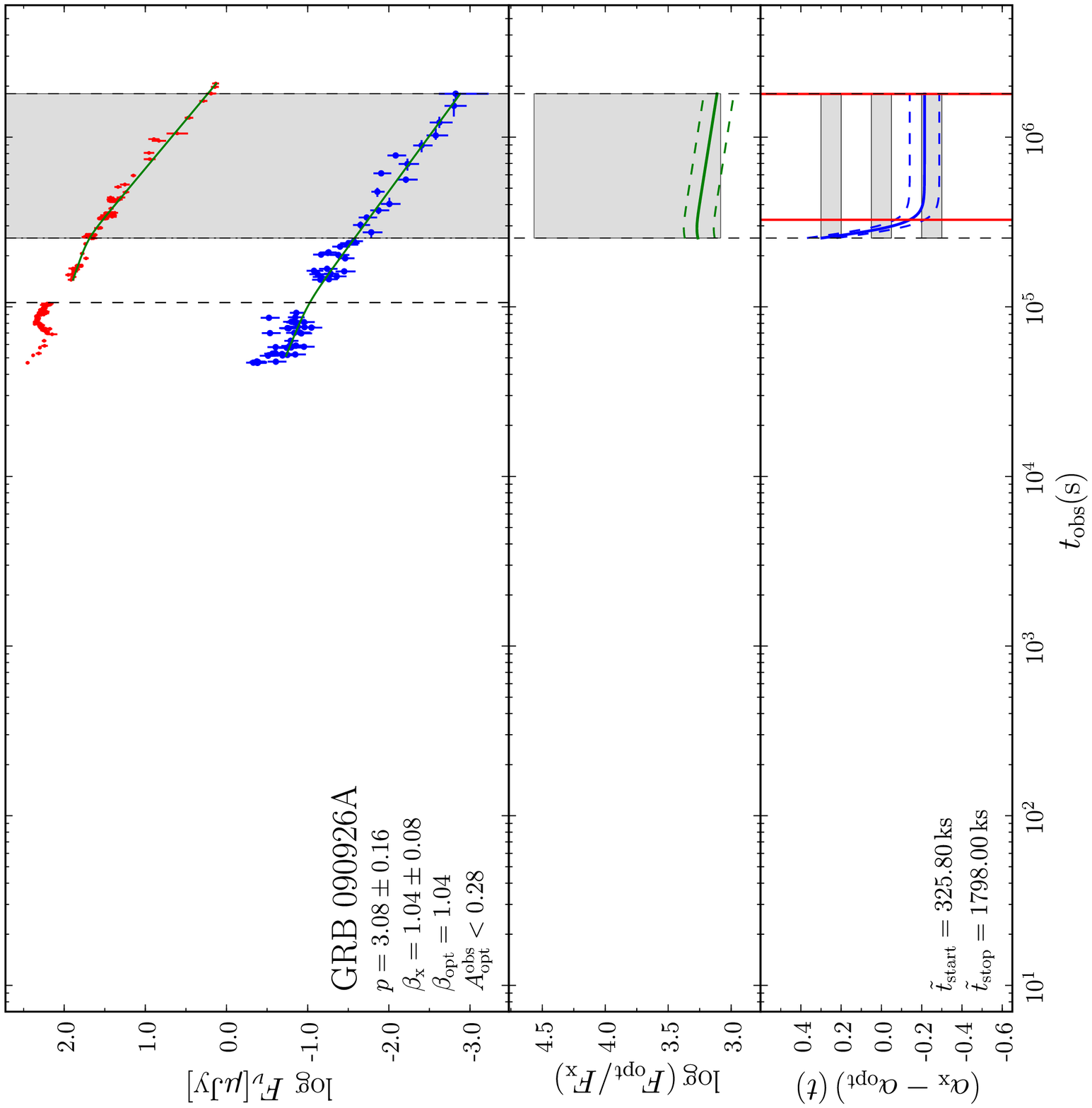}
\hfill\center{Fig. \ref{fig:sample} ---  continued}
\end{figure*}